\documentclass[fleqn,usenatbib]{mnras}
\UseRawInputEncoding
\usepackage{xcolor}
\usepackage{newtxtext,newtxmath}
 
\usepackage[T1]{fontenc}

\usepackage[normalem]{ulem}

\DeclareRobustCommand{\VAN}[3]{#2}
\let\VANthebibliography\thebibliography
\def\thebibliography{\DeclareRobustCommand{\VAN}[3]{##3}\VANthebibliography}

\usepackage{graphicx}	
\usepackage{amssymb}
\usepackage{amsmath}	
\usepackage{threeparttable}

\definecolor{address}{rgb}{0.36,0.54,0.66}
\definecolor{red}{rgb}{0.8,0.,0.}

\newcommand{\eagle}{\textsc{\large eagle}{ }}
\newcommand{\tng}{\textsc{\large illustristng}{ }}

\newcommand{\simba}{\textsc{\large simba}{ }}

\usepackage{orcidlink}

\graphicspath{{Figures/}}

\title[Dark matter properties from stellar light]{Inferring the dark matter distribution of massive galaxy clusters from deep optical observations: insights from the TNG300 simulation}
\author[A. Manuwal et al.]{
\parbox{13.5cm}{
Aditya Manuwal$^{\orcidlink{0000-0003-2893-2793}}$\thanks{E-mail:adi.manuwal@astro.unam.mx},$^1$
Vladimir Avila-Reese$^{\orcidlink{0000-0002-3461-2342}}$,$^1$
Daniel Montenegro-Taborda$^{\orcidlink{0000-0002-5428-9984}}$,$^2$
Vicente Rodriguez-Gomez$^{\orcidlink{0000-0002-9495-0079}\,2}$ and
Bernardo Cervantes Sodi$^{\orcidlink{0000-0002-2897-9121}\,2}$
}
\vspace{0.3cm}
\\
$^{1}$Universidad Nacional Aut\'onoma de M\'exico, Instituto de Astronom\'{\i}a, A.P. 70-264, 04510 CDMX, M\'exico\\
$^{2}$Instituto de Radioastronom\'{\i}a y Astrof\'{\i}sica, Universidad Nacional Aut\'onoma de M\'exico, A.P. 72-3, 58089 Morelia, M\'exico
}

\date{Accepted 2025 October 02. Received 2025 August 30; in original form 2025 May 09}

\pubyear{2025}

\begin{document}
\label{firstpage}
\pagerange{\pageref{firstpage}--\pageref{lastpage}}
\maketitle

\begin{abstract}
Extragalactic stars within galaxy clusters contribute to the intracluster light (ICL), which is 
thought to be a promising tracer of the underlying dark matter (DM) distribution. In this study, we employ the TNG300 simulation to investigate the prospect of recovering the dark matter distribution of galaxy clusters from deep, wide-field optical images. For this, we generate mock observations of 40 massive clusters ($M_{200}\gtrsim 10^{14.5}\,{\rm M}_\odot$) at $z=0.06$ for the $g'$ band of the Wendelstein Wide-Field Imager (WWFI), and isolate the emission from the brightest cluster galaxy (BCG) and the ICL by masking the satellite galaxies, following observational procedures. By comparing $\Sigma_{\rm BCG+ICL}$ profiles from these images against $\Sigma_{\rm DM}$ profiles for the central subhaloes, we find that $\Sigma_{\rm cen-DM}/\Sigma_{\rm BCG+ICL}$ exhibits a quasi-linear scaling relation in log space with the normalised distance $r/R_{\Delta}$, for both $R_{\Delta}=R_{200}$ and $R_{500}$. The scatter in the scaling is predominantly stochastic, showing a weak dependence on formation time and dynamical state. We  recover the DM concentration and mass within $\approx 23$ and $\approx 15$ per cent of their true values (for $R_{200}$), respectively, and with $\approx 3$ per cent larger uncertainties for $R_{500}$. Alternatively, we find that the concentration can be estimated using the BCG+ICL fraction, the central's DM mass using the BCG+ICL flux, and the total DM mass using the bolometric flux. These results demonstrate the feasibility of deriving dark matter characteristics of galaxy clusters to be observed with facilities like the Vera C. Rubin Observatory in the near future. 
\end{abstract}

\begin{keywords}
galaxies: clusters: general -- dark matter -- hydrodynamics -- methods: data analysis -- methods: numerical
\end{keywords}



\section{Introduction}
Dark matter (DM) comprises $\approx 85$ per cent of the matter in the Universe \citep{Planck2020} 
and portrays a nature broadly aligning with the cold DM (CDM) hypothesis \citep{Blumenthal1984,Davis1985,Percival2001}. The concordance $\Lambda$CDM model states that the formation of structures has its roots in the primordial overdensities in DM set by quantum fluctuations during its early history. The overdensities amplified over time owing to the accretion of surrounding material until 
they collapsed and lead to bound structures \citep[see
the review by][]{Zavala2019}. These are formally referred to as `haloes' and serve as sites for galaxy formation, which occurs when they accumulate cold gas in their central regions \citep[for details, see][]{Mo2010,Almeida2014}. Haloes grow further by smooth accretion of matter or mergers with other haloes, where the latter aids in gaining satellite galaxies and the formation of groups.

Galaxy clusters are associated to the most-massive haloes ($\gtrsim 10^{14}\,{\rm M}_\odot$) resulting from this hierarchical assembly, and have provided one of the earliest
support in favour of DM's existence \citep{Zwicky1933,Smith1936,Zwicky1937} and its collisionless dynamics \citep{Markevitch2004,Clowe2006} -- that is, 
the interaction cross-section is small enough so that the two-body interactions between 
the particles do not affect the overall phase structure. The distribution of cluster mass and its 
evolution is a useful probe of the physics of gravitational collapse and the underlying cosmology 
\citep{Vikhlinin2009,Allen2011}. Clusters also serve as important cosmological laboratories to test theories of 
galaxy formation within a given cosmological framework, owing to the peculiar phenomena that occur within these extreme environments \citep[see][]{Kravtsov2012}. The strong tidal field near the cluster's centre causes 
stripping of satellites on their pericentric approaches \citep{Gnedin2003,Coenda2009,Ramos2015,Smith2016}. Likewise, the high orbital velocities are conducive for ram pressure stripping caused by interactions 
between the galaxy and the intrahalo gas \citep{Abadi1999,Domainko2006,Ebeling2014,Steinhauser2016}.
Among other factors, the efficiencies of these phenomena are dependent on the mass and its distribution within the cluster \citep{Read2006,Singh2019,Roberts2021}, and this warrants effective ways of its characterisation.

Naturally, one has to rely on indirect methods for this exercise. The most accurate way is gravitational
lensing, which measures distortions in images of background sources caused by the space-time curvature in and around the cluster (see \citealt{Natarajan2024} and the references therein). This technique allows one to
map both the radially-averaged and two-dimensional distributions, and even has the potential to
identify the substructures devoid of galaxies \citep{Ghosh2021,Wagner2023}. Alternatively,
cluster mass can be derived by assuming hydrostatic equilibrium for the intracluster gas, and using X-ray or Sunyaev-Zel'dovich (SZ) measurements to infer gravitational potential from pressure and density profiles \citep{Bartalucci2018,Ettori2019}. One can also utilise the projected phase-space of satellites and assume virial theorem to obtain a dynamical mass estimate \citep[e.g.][]{Mamon2013}. 


These methods are highly useful, but cumbersome to implement. Moreover, they recover the \textit{dynamical} (total) mass distribution in the evolved clusters, i.e., taking into account the halo (main subhalo) and the subhaloes.  However, if the goal is to infer or constrain from observations the properties of the ``cosmological'' DM haloes and subhaloes that emerged from the density perturbation field, it is important to make the distinction between the two. The galaxy--halo connection -- widely used in theoretical (semi-analytical models) and statistical (semi-empirical modeling) approaches -- technically addresses how central/satellite galaxies populate haloes/subhaloes from N-body cosmological simulations. Therefore, to validate this connection for observed central galaxies, it is necessary to infer the DM characteristics of the \textit{main} subhalo, such as its radial distribution and concentration. Note that in the case of galaxy clusters, the main subhalo is linked to the central galaxy and the diffuse stellar component, whereas satellite galaxies populate the subhaloes. This requires novel, inexpensive methods for explicitly deriving the DM distribution within the \textit{main} subhaloes of clusters, which is the primary focus of this work.


A physically-motivated approach for devising such a method is to utilise a cluster component that occupies similar phase-space as DM, and is also straight-forward to observe. Stars do exhibit near-collisionless dynamics like DM, but a bulk of them are trapped within galaxies, which
are biased tracers of the DM halo. However, it is now well-established that some fraction of stellar light from clusters is diffuse (e.g. surface brightness, $\mu \gtrsim 27~{\rm mag}\,{\rm arcsec}^{-2}$) and originates in the intergalactic space \citep[see the review by][]{Montes2022}. This diffuse emission, termed as the intracluster light (ICL), 
is already in place at $z\gtrsim 1$ \citep{Ko2018,Joo2023} and builds up via liberation of stars during tidal stripping of massive satellite galaxies, violent relaxation during mergers with the
brightest cluster galaxy (BCG), \textit{in situ} star formation in the intracluster medium, and preprocessing of member galaxies in groups before they fall into a cluster \citep[see][for a review]{Contini2021}. The relative dominance of these channels is still debated and is an active area of research \citep[e.g.,][]{Chun2023,Montenegro2023,Ahvazi2024,Brown2024,Contini2024,Bilata-Woldeyes2025,Montenegro-Taborda2025}.

In addition to the collisionless aspect, what makes ICL a promising DM tracer is its smooth distribution 
and large extent, reaching hundreds of kpc \citep[e.g.][]{Gonzalez2005,Zibetti2005,Krick2007,Toledo2011}. This suggests the dynamics of ICL stars is predominantly governed by the global cluster potential rather than individual substructures, and that it can be used to infer the 
DM distribution across the whole halo. The validity of this idea has been demonstrated for observed clusters in terms of the differential radial profile \citep{Sampaio2021} and two-dimensional distribution \citep{Montes2019,Kluge2021,Diego2023}.

One can investigate this in a more thorough and straight-forward manner with simulations that explicitly model both DM and stars and evolve them self-consistently. There has been a recent surge of work in this regard due to rising interest and improvements in computational techniques/power 
\citep{Alonso2020,Sampaio2021,Shin2022,Yoo2022,Reina2023,Contreras2024,Yoo2024,Butler2025}. 
A broad consensus emerging from these studies is: 
\begin{enumerate}
    \item the close correspondence between the two-dimensional distributions of ICL and DM,
    \item greater similarity for relaxed clusters,
    \item the radial profile of ICL being steeper than that of DM, and
    \item the ICL and DM surface density profiles follow a power-law relationa as a function of cluster-centric radius.
\end{enumerate}
This is despite the fact that the simulations used in these works vary in physical models, computation scheme, and ICL definitions.

Additionally, there are indications that the combined light from BCG and ICL is also a good tracer
of DM, and should, in fact, be preferred over ICL only. \citet{Yoo2022} quantified differences in projected distributions of DM and potential tracers (all stars, galaxies, and BCG+ICL) in the Galaxy Replacement Technique simulation \citep{Chun2022} using their novel weighted-overlap-coefficient measure. They showed that BCG+ICL is a better tracer than galaxies or the total stellar component, and the degree of similarity with DM is greater for dynamically relaxed clusters. This has also been confirmed for the Horizon Run 5 simulation \citep{Lee2021} by \citet{Yoo2024}. Furthermore, \citet{Yoo2024} showed that, just like ICL \citep{Alonso2020,Reina2023,Contreras2024}, the BCG+ICL profile can be used to obtain the DM profile using a scaling relation with radius. These results are particularly helpful from an observational perspective, because what appears as BCG in an optical image also contains light from ICL stars located between the BCG and the observer, and it is non-trivial to isolate the ICL out of this emission. Moreover, a detailed comparison of various characterisation methods for deep optical observations indicates that the BCG+ICL measurement is more robust than ICL \citep[see][]{Brough2024}.

Note that a relation between diffuse light and DM distribution derived from simulations 
is directly applicable to observations only if one properly accounts for factors that are encountered in a realistic scenario. For one, relations based on stellar \textit{mass} distributions \citep[e.g.,][]{Alonso2020,Yoo2024} implicitly assume a constant mass-to-light ratio throughout the halo,
which is not necessarily true. Converting these to light maps can circumvent this issue and involves computing magnitudes for a filter of choice from each stellar particle using a stellar population synthesis model \citep[as done by][for example]{Reina2023}. However, there are additional effects from the instrumental point-spread-function (PSF) and various sources of noise that are intrinsic to the image, like sky background \citep[see][]{Kluge2020}. One could ignore them for mock observations assuming a scenario where the effects have already been removed perfectly from real observations before analysis, but this strictly holds only in an ideal situation. A more convenient approach is to forward-model these aspects in the simulated image, especially for the ease of application to observations.

Apart from just for the sake of realism, incorporating these effects is important to facilitate direct
applicability of results to data products from the imminent, wide-area, deep optical surveys that are poised to yield observations of clusters down to unprecedented sensitivities. The Legacy Survey of Space and Time (LSST; \citealt{Ivezic2019}) planned for the Vera C. Rubin Observatory will observe $20000\,{\rm deg}^2$ of the southern sky for 10 yr and reach $\mu \gtrsim 30.5\,{\rm mag}\,{\rm arcsec}^{-2}$ \citep{Brough2020}. This will provide the ideal data set for low surface brightness science in the coming years. Likewise, Euclid \citep{Laureijs2011} is expected to observe nearly one-third of the sky ($15000\,{\rm deg}^2$) with a limiting brightness of $\approx 29.5\,{\rm mag}\,{\rm arcsec}^{-2}$.

With this as our motivation, we investigate the prospects of inferring DM properties of massive clusters from deep optical images using the largest volume run of the \tng suite of magneto-hydrodynamic simulations \citep{Weinberger2017,Pillepich2018}. We base our results on mock observations of 40 clusters at $z\approx 0.06$ for the $g'$ band of Wendelstein Wide-Field Imager (WWFI\footnote{\url{https://www.usm.uni-muenchen.de/wendelstein/htdocs/wwfi.html}}; \citealt{Kosyra2014}) onboard the 2-m Fraunhofer telescope at the Wendelstein Observatory. 
Broadly, our work uses realistic synthetic images and DM maps to derive relationships between photometric measurements and DM properties of galaxy clusters, the likes of which can be applied directly to suitable observations -- for example, those by \citet{Kluge2020} obtained with the WWFI for a sample of 170 clusters with a median redshift of $z\approx0.06$.

The paper is organised as follows. In Section~\ref{methods}, we describe the simulation and our sample (Section~\ref{sim}), the generation of mock images (Section~\ref{imagegen}), and their analysis (Section~\ref{analysis}). The results from this work are described and discussed in Sections~\ref{uniscale}
and \ref{global}. Section~\ref{uniscale} presents the universal scaling relation between DM surface density 
profile for the central subhalo and the optical BCG+ICL surface brightness profile based on masked images.
This forms one of the main results from this work. The section also explores the recovery of DM parameters from the scaling, and discusses the factors modulating its scatter. In Section~\ref{global}, we explore the prospective of using various global photometry measurements
to infer the DM parameters for the central subhalo, along with the total DM mass. Additionally, 
we compare the predictions for central subhalo's DM parameters from the scaling method in Section~\ref{uniscale} against those from global photometry, and provide recommendations for their applications. In Section~\ref{caveats}, we explore the caveats that may limit direct application of our results
to real clusters, and provide recommendations to correct for them. Finally, we summarise our analysis, main findings, and conclusions in Section~\ref{conclude}. All the lengths are given in physical/proper units, unless specified otherwise.

\section{Methods}\label{methods}
\subsection{The simulation}\label{sim}
\tng is a suite of cosmological magneto-hydrodynamical simulations run with the moving-mesh code \textsc{\large arepo} \citep{Springel2010}, and a comprehensive model for baryonic physics that realistically follows the formation and evolution of galaxies across cosmic time \citep{Weinberger2017,Pillepich2018}. Each simulation treats the coupled evolution of DM, gas, stars, and supermassive black holes (SMBHs) from $z=127$ to $z=0$
assuming the \citet{Planck2016} cosmology -- i.e. $\Omega_{\rm \Lambda,0}=0.6911,\Omega_{\rm m,0}=0.3089,\Omega_{\rm b,0}=0.0486,\sigma_{8}=0.8159,n_s = 0.9667$ and $h=0.6774$. It incorporates the physics for radiative cooling and heating with an evolving ultraviolet background; star formation and evolution; chemical enrichment associated with supernovae and asymptotic giant branch stars; galactic-scale outflows driven by stellar feedback; the formation, merging and accretion of SMBHs; and two black hole (BH) feedback modes dependent on the BH accretion rate.

Since these simulations cannot resolve small-scale processes (like star formation), they are
implemented via subgrid models tuned to reproduce various observed galaxy/halo properties and statistics.
The simulations are nevertheless consistent in several other properties that were not employed for the calibration, thereby demonstrating their predictive power. This includes the aspects that are particularly relevant for our study, like the ram-pressure stripped, jellyfish galaxies \citep{Yun2019}; scaling relation between cluster mass, Sunyaev-Zel'dovich parameter, total radio power and X-ray luminosity \citep{Marinacci2018}; statistics of cool core clusters \citep{Barnes2018}; the distribution of metals in the intracluster medium \citep{Vogelsberger2018}; and the velocity dispersion of the BCG stars and the satellite galaxies \citep{Sohn2022}.

In this work, we employ the $z\approx 0.06$ snapshot of the highest resolution version of the largest volume runs, `TNG300-1' (\citealt{Nelson2019}; hereafter referred to as TNG300). It includes $2\times 2500^3$ resolution elements in a $302.6^3\,{\rm Mpc}^3$ comoving volume with the (\textit{initial}) baryonic and dark matter mass of $1.1\times 10^7\,{\rm M}_\odot$ and $5.9\times 10^7\,{\rm M_\odot}$, respectively. The large volume enables ample sampling
of clusters, making it the most suitable run for our science. The gas in the simulation is discretised using Voronoi tesselation, resulting in a highly adaptive spatial-resolution that increases for denser regions; the minimum cell size being $370$ pc. The gravitational softening length of the collisionless elements (DM and stars) is $1.48$ kpc.

The baryonic structures are identified in a two-step process. First, the Friends-of-Friends (FoF) algorithm \citep{Davis1985} is applied on DM particles to detect DM haloes using the linking length of $b = 0.2$ times the mean inter-particle spacing. Each baryonic element is assigned to the halo with the DM particle nearest to it. Then, gravitationally-bound substructures (or subhaloes) are discerned within each halo via \textsc{\large subfind} \citep{Springel2001,Dolag2009} using all the particle species. The most massive subhalo is called `central' while the rest are called `satellites'. The centre of each subhalo is assumed as the location of the minimum gravitational potential, and the centre of each parent/group halo corresponds to that of the central.

Note that the terms `central' and `satellite' in \textsc{\large subfind} outputs refer to \textit{all} the matter (DM+baryons) in these substructures. Hereafter, we use the term central and satellite \textit{subhaloes} for the respective DM components, `BCG+ICL' for the central's stellar component, and `satellite galaxy' for the satellite's stellar component.

The mass of each group, $M_{\Delta}$, is evaluated based on the total matter enclosed within a radius, $R_{\Delta}$, where $R_{\Delta}$ is taken as a proxy for the halo radius and corresponds to the mean density of $\Delta$ times the critical density of the universe, $\rho_{{\rm crit}} = 3H(z)^2/{8\pi G}$ [$H(z)$ is the Hubble parameter at the redshift and $G$ is the gravitational constant]. The results in this paper are based on both $\Delta=200$ and $\Delta=500$, where the former is a physically motivated measure and the latter is more reminiscent of observational estimates, usually restricted to smaller radii due to detection limits \citep[e.g.][]{Pratt2016,Biviano2017,Bartalucci2018}. In some locations, we have used 
the masses based only on the DM content, and if so, we have explicitly mentioned this in the text and have distinguished them through appropriate notations.

An extensive mapping of the stellar distribution within a cluster requires a high signal-to-noise data out to $r\approx R_{200}$. This can be achieved best for massive clusters, as they are supposed to contain significant ICL mass. In this work, we focus our analysis on the 40 clusters with $M_{200}\gtrsim 10^{14.5}\,{\rm M}_\odot$ in the $z\approx0.06$ snapshot of TNG300. The details regarding image production and analysis are described in the subsection below.

\subsection{Mock observations}\label{imagegen}
\begin{figure*}
    \centering
    \includegraphics[width=2.07\columnwidth]{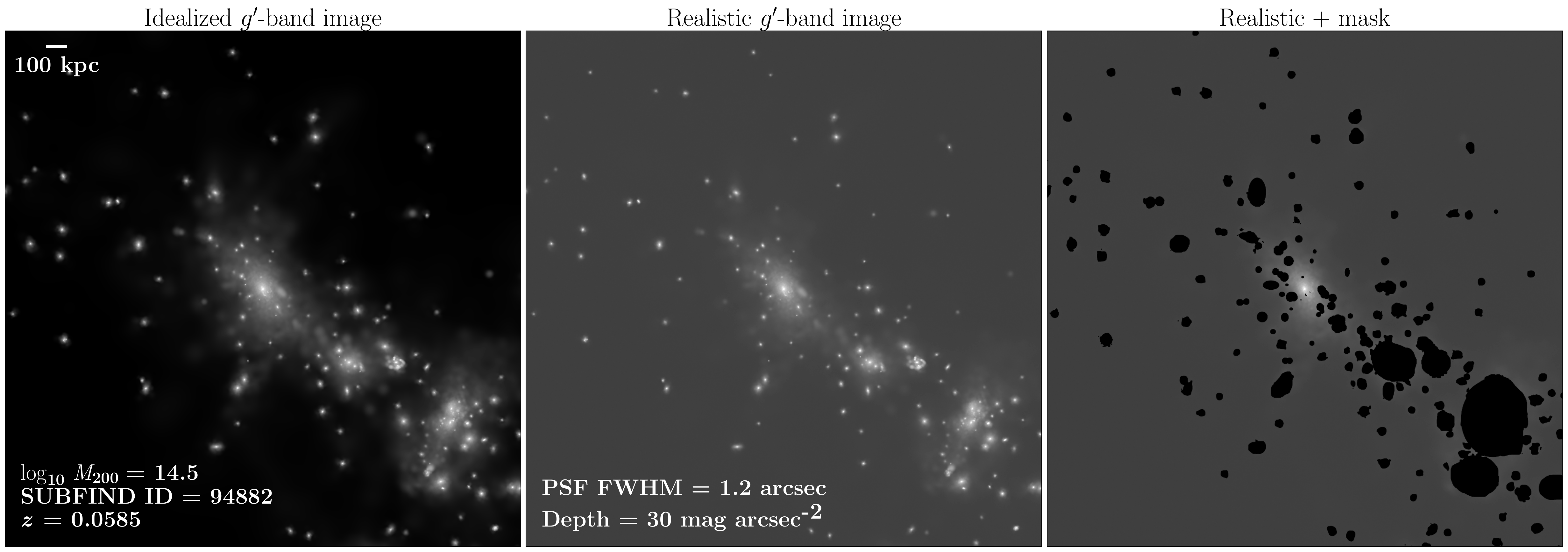}
    \caption{An illustration of the synthetic WWFI-like images generated for the TNG300 clusters (see Section~\ref{imagegen} for details). The images shown here correspond to the cluster with the \textsc{\large subfind} ID 94882. The idealised image (without incorporating observational factors) for the $g'$-band is shown in the left panel. The middle panel shows the realistic image
    obtained after simulating the impact of the PSF, background noise, and shot noise. The right panel
    shows the realistic image with masked satellite galaxies used to characterise the BCG+ICL component.}
    \label{illust}
\end{figure*}

Our methodology for generating synthetic optical images for the clusters was presented in \citet{Montenegro-Taborda2025b} and can be broadly summarised in two stages, similar to those delineated in \citet{Rodriguez2019}: a) producing the idealised images based only on the stellar emission, and b) post-processing these images to implement the effects from PSF and noise. We use here the same sample of 40 cluster mock images from \citet{Montenegro-Taborda2025b}. The goal in that paper was to emulate the cluster observations made by \citet{Kluge2020} with the WWFI at the Wendelstein Observatory. All the observational parameters are set to match these observations. 
An illustration of the resultant images is provided in Fig.~\ref{illust}.

\subsubsection{Idealised images}
In principle, the most comprehensive approach for this stage is to compute stellar emission using radiative transfer to account for dust-radiation interactions. One can employ publicly-accessible codes for this purpose \citep[e.g.][]{Baes2011,Robitaille2011}, but these are overly expensive for systems with negligible amounts of star-forming gas and dust content\footnote{This study focusses on BCG and ICL,
both of which have meager dust contents. Dust attenuation can be
significant for satellite galaxies, but these are removed in our analysis (see Section~\ref{isolate}). The magnitudes used in Section~\ref{scatter} have been obtained after accounting for dust effects.} Instead, \citet{Montenegro-Taborda2025b} employed an image generation pipeline\footnote{\url{https://github.com/vrodgom/galaxev_pipeline}} based on the \textsc{\large galaxev} stellar population synthesis code \citep{Bruzual2003,Bruzual2011} whose results are practically indistinguishable to those from radiative transfer codes, but reduces the computation time by at least two orders of magnitude [compared to the \textsc{\large skirt} (\citealt{Baes2011}) pipeline, for example; see \citealt{Rodriguez2019}].

In this framework, the emission of each stellar particle is computed using 
simple stellar population (SSP) models built on the `Padova 1994' evolutionary tracks and a \citet{Chabrier2003} initial mass function. These models contain the rest-frame luminosity per unit wavelength of an SSP as a function of wavelength, metallicity, and age. These luminosities were converted to observer-frame values by assuming that the source is located at $z= 0.06$ and accounting for the cosmology. Then, the magnitudes were estimated
based on the response function for the $g'$ filter (obtained from the SVO Filter Profile Service\footnote{\url{http://svo2.cab.inta-csic.es/theory/fps/}}; \citealt{Rodrigo2020}) and stored in a grid that can be interpolated to derive the apparent magnitude of any stellar particle for its initial mass, age, and metallicity. 

To create the image, \citet{Montenegro-Taborda2025b} considered the image resolution of 0.2~arcsec per pixel, assumed the line-of-sight perpendicular to the $x$-$y$ plane of the simulation, centred 
the field-of-view at the cluster's centre, and extended it out to a radial extent of $R_{200}$. Given that each stellar particle in the simulation represents a SSP, the density was smoothed by convolving with a `standard' spline kernel described in equation (1) of \citet{Rodriguez2019}, where the adaptive smoothing scale was chosen as the distance to the 64th nearest stellar particle [similar to the approach in \citet{Torrey2015}]. Finally, the fluxes were added from all the smoothed stellar particles in each pixel. By default, these flux values are
in analog-to-digital units per second (${\rm ADU}\,{\rm s}^{-1}$), which are scaled such that the zero-point magnitude is 30 $g'$ mag. An example of such an idealised image\footnote{Note that the idealised images are \textit{not} equivalent to projected stellar mass maps because mass-to-light ratio varies across the cluster (Appendix~\ref{masstolumvsrad}).} can be seen in the left panel of Fig.~\ref{illust}.

\subsubsection{Applying realism}
In a realistic setting, the observed image includes the influence from telescope optics, and also from atmospheric turbulence if it is a ground-based instrument. \citet{Montenegro-Taborda2025b} incorporated these effects by convolving the idealised images with a symmetric Gaussian PSF with full width at half-maximum (FWHM) as the median seeing for the $g'$ band from the 2 meter telescope at Wendelstein Observatory, i.e., ${\rm FWHM}=1.2~{\rm arcsec}$. 

For modeling the sky background noise, it was assumed that the pixels associated with galactic flux are surrounded by `sky' pixels whose values follow a Gaussian distribution. The average value of the standard deviation of these pixels is $\sigma_{\rm sky}=2~{\rm ADU}\,{\rm s}^{-1}\,{\rm pixel}^{-1}$ in the observations (determined by \citealt{Montenegro-Taborda2025b} via sigma clipping). The noise to each pixel was added in a synthetic image by drawing a random value from the Gaussian distribution with this $\sigma_{\rm sky}$. 

There can be additional fluctuations in the pixel value due to discretisation, usually termed `shot noise'. We can implement this by determining the exact number of electrons corresponding to a pixel value, which requires knowledge of the instrumental gain. Given
that $1e^{-}{\rm s}^{-1}$ corresponds to 25.4 $g'$ mag \citep{Kosyra2014} and the zero-point is defined as 30 mag, we calculate the gain as
\begin{equation}
G = 10^{0.4(25.4 - 30.0)}, 
\end{equation}
and use it to convert the pixel flux in ${\rm ADU}\,{\rm s}^{-1}$ to $e^{-}{\rm s}^{-1}$. We then multiply this by the exposure time of 3120~seconds (the median value in the observations) to determine the electron count, and randomly sample from the Poisson distribution with the expected number of events as the count. The final image looks
something like the middle panel of Fig.~\ref{illust}, which is visibly noisier than the idealised version in the panel to its left.

\subsection{Isolating the combined emission from BCG and ICL}\label{isolate}
Segregating the stellar distribution into BCG, ICL, and satellite galaxies is an arduous task due to their overlap in phase-space \citep[e.g.,][]{Proctor2024}, and is especially difficult in observational data due to the lack of full three-dimensional information. Therefore, there are a myriad of methods that have
been suggested for extracting these components from observations (see e.g., \citealt{Brough2024,Montenegro-Taborda2025}, and more references therein). These provide disparate results for ICL, but nevertheless demonstrate remarkable consistency for the combined BCG+ICL emission \citep[e.g.][]{Brough2024}. We also note that using BCG along with ICL avoids the assumptions implicit in modelling these components to segregate their respective contributions. Hence, we do not attempt to separate the two.

We extract the BCG+ICL flux for each cluster using an observationally motivated, non-parameteric approach based on image masking\footnote{Ideally, one would use the central's
stellar emission to capture the `true' BCG+ICL, but this is non-trivial to recover from
observations in practice. Furthermore, one cannot simply replace the masked image with the central's 
image; the profiles for our clusters can vary by $\lesssim 40$ per cent across radii (see Appendix~\ref{maskedvscencomp}).}. Specifically, \citet{Montenegro-Taborda2025b} apply the automated masking algorithm in \citet{Kluge2020}. Although this suffices for satellite galaxies that appear as distinct objects, the masks may fail to capture peculiar features like tidal tails. In such cases,
we manually adjust the masks as needed. An example of this masking is portrayed in the right panel of Fig.~\ref{illust}. All the measurements for BCG+ICL in this work are based exclusively on the
unmasked pixels.

\subsection{Surface brightness/density profiles}\label{analysis}
To compute the BCG+ICL surface brightness ($\Sigma_{\rm BCG+ICL}$) profile, we divide the masked optical image into concentric circular annuli around the
image centre with evenly-spaced logarithmic radii ranging from $10^{-2.2}$ arcmin to $10^{1.5}$ arcmin and seperated by $0.1$~dex. This is implemented through the modified version of routines in the \textsc{\large pyproffit} package \citep[described in][]{Eckert2020}. The radial span is chosen to encompass radii up to $R_{200}$ for all the clusters. For each annulus, we compute the surface brightness by summing up the flux from the unmasked pixels, taking the median value, and dividing it by the area of the annulus. For most of the clusters, the surface brightness profiles obtained with circular apertures are very similar to those obtained with elliptical apertures following the isophotal contours \citep{Montenegro-Taborda2025b}.
We compute the DM surface mass density ($\Sigma_{\rm cen-DM}$) profile in a similar way, using
the projected DM mass map of the \textit{central (main) subhalo} and assuming the same resolution and field of view as the corresponding optical image.

We also computed DM surface densities using the projected 
\textit{total} DM mass maps and masking them with the masks used in the surface brightness maps ($\Sigma_{\rm DM,\,masked}$). The differences between $\Sigma_{\rm cen-DM}(r)$ and $\Sigma_{\rm DM,\,masked}(r)$ are generally very small, but at large cluster-centric radii the profiles from the masked maps can be up to 15 per cent higher than those of the central subhaloes; see Fig.~\ref{maskedvscen} and a discussion of why this happens in the Appendix~\ref{maskedvscencomp}.

\section{The universal $\Sigma_{\rm cen-DM}$--$\Sigma_{\rm BCG+ICL}$ scaling relation}\label{uniscale}
Upon examining individual clusters in our sample, we find that the $\Sigma_{\rm BCG+ICL}$ profiles are always steeper (more concentrated) than the $\Sigma_{\rm cen-DM}$ profiles. This echoes previous results from both observations of galaxy clusters \citep[e.g.][]{Zhang2019,Diego2023} and hydrodynamical simulations \citep{Pillepich2018,Alonso2020,Sampaio2021,Shin2022,Yoo2024}. The precise physics behind this trend has not been studied yet, but it likely originates from the fact that, as a satellite enters a cluster, its DM gets stripped first due to its (typically) larger scale, and most of the stellar stripping occurs later after the satellite's orbit has degraded due to dynamical friction \citep[e.g., see][]{Smith2016,Engler2021,Montero2024}.

\begin{figure*}
    \centering
    \includegraphics[width=2.085\columnwidth]{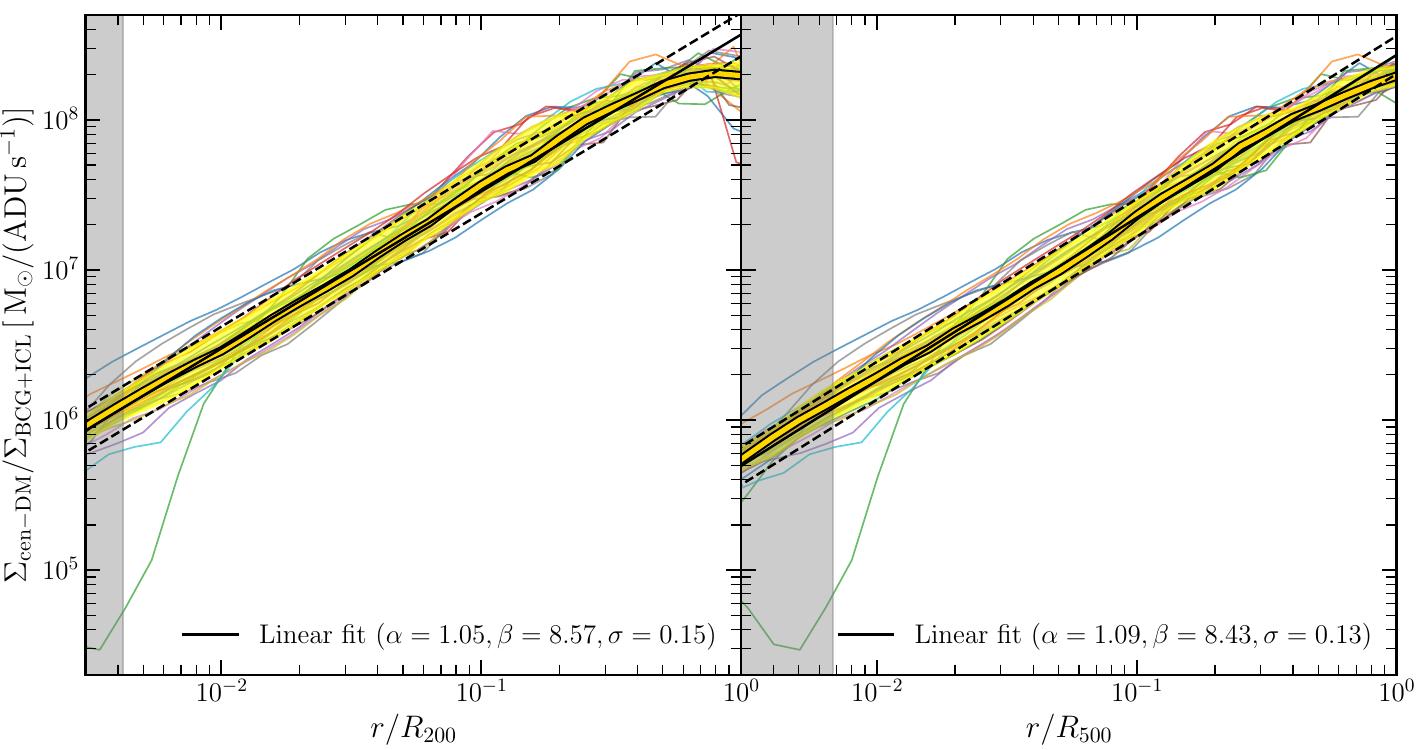}
    \caption{The scaling relation between the intrinsic central subhalo's DM surface density profile and the `masked' BCG+ICL surface brightness profile for TNG300 clusters. The left and right panels show $\Sigma_{\rm cen-DM}/\Sigma_{\rm BCG+ICL}$ against cluster-centric radii normalised by $R_{200}$ and $R_{500}$, respectively. The $\Sigma_{\rm BCG+ICL}$ profile is derived from the optical image after masking the emission from satellite galaxies, and the DM profile is calculated from the DM map for the central subhalo. Each colour denotes a specific cluster (Section~\ref{analysis}). The grey shaded region in the left shows the regime where the profiles are unreliable
    due to numerical heating driven by two-body scattering of particles (see the text). The yellow curve shows the median relation, and the corresponding yellow shaded region spans 16th to 84th percentiles. The solid black line is the maximum-likelihood linear fit to the profiles beyond the shaded region, and the
    dashed lines show 1-$\sigma$ scatter from the fit along $y$-axis. The slope ($\alpha$),
    intercept ($\beta$), and the $\sigma$ are mentioned in the bottom-right corner of each panel.}
    \label{scaling}
\end{figure*}

More importantly, we find that the ratio between $\Sigma_{\rm cen-DM}$ and $\Sigma_{\rm BCG+ICL}$ varies with radius broadly by the same degree for different clusters in our sample. To assess this self-similarity, we plot $\Sigma_{\rm cen-DM}/\Sigma_{\rm BCG+ICL}$ against the normalised cluster-centric radius ($r/R_\Delta$)
in Fig.~\ref{scaling}. The left and right panels show the profiles with radii normalised by $R_{200}$ and $R_{500}$, respectively. Each curve corresponds to a given cluster and is represented with a specific colour. Here we have excluded the two clusters (IDs 31816 and 103007) from this plot that exhibit no surface brightness at the innermost radii due to masking of one or more satellite galaxies overlapping with the BCG in projection. The median relation based on the 38 clusters is shown as the solid yellow curve, and the 16th to 84th percentiles are shown using the yellow shaded region around the
median.

In addition, we note that the limited mass resolution of DM particles results in a
coarse-grained representation of an otherwise smooth gravitational potential, causing artificial
potential fluctuations within the halo. The two-body scattering between particles can modify their kinematics and spatial distribution, particularly in the dense regions where the relaxation timescale is smaller than the Hubble time \citep[see][]{Ludlow2019}. Stellar orbits can also get modified via encounters with DM particles, but this effect is negligible within massive haloes at the scales of galaxy clusters \citep{Ludlow2021,Ludlow2023,Wilkinson2023}. This means that the DM profiles below a certain radius are unreliable. We use equation~(19) in \citet{Ludlow2019} to determine this convergence radius,
\begin{equation}\label{convrad}
r_{\rm conv} = 0.77\left(\frac{3\Omega_{\rm dm,0}}{800\pi}\right)^{1/3}l,    
\end{equation}
where $l$ is the mean inter-particle spacing,
\begin{equation}
l = \frac{La}{N_{\rm p}^{1/3}},    
\end{equation}
for $N_{\rm p}$ particles in a box of side-length $L$ at a scale factor $a$. This yields a $r_{\rm conv}=5.9$~kpc (or 0.09 arcmin) for our snapshot, which is $\approx 4$ times the gravitational softening length. The grey shaded region in Fig.~\ref{scaling} spans the (projected) radii below the \textit{maximum} $r_{\rm conv}/R_{200}$ (or $r_{\rm conv}/R_{500}$) among all the 38 haloes considered here.

The figure reveals an explicit universal scaling relation between the two profiles for massive galaxy clusters. This median relation\footnote{The median values and the 16th-84th percentiles of $\Sigma_{\rm cen-DM}/\Sigma_{\rm BCG+ICL}$ as a function of $r/R_{200}$ and $r/R_{500}$ (see Fig.~\ref{scaling}) are provided online as supplementary material.} is almost linear in the log-log space regardless of the cluster radius choice. We parameterise it as a linear relation given by
\begin{equation}\label{fit}
\log_{10}(\Sigma_{\rm cen-DM}/\Sigma_{\rm BCG+ICL}) = \alpha\log_{10}(r/R_\Delta)+\beta \pm \sigma,    
\end{equation}
with $\sigma$ being the uncertainty. To obtain the fitting parameters, we consider the profiles outside the grey area (i.e. $r\gtrsim 0.004R_{200}$ for the left panel and $r\gtrsim0.007R_{500}$ for the right panel), and maximise the Gaussian log-likelihood defined in equation~(13) of \citet{Robotham2015}\footnote{All the linear fits shown in this work have been derived through
this approach.}:
\begin{equation}
\ln{\mathcal{L}} = 0.5\sum_{i=1}^{N}\left[\ln{\frac{\alpha^2+1}{\sigma^2}-\left(\frac{\alpha x_i+\beta-y_i}{\sigma}\right)^2} \right],    
\end{equation}
where $x_i = \log_{10}(r/R_\Delta)_i$, $y_i=\log_{10}(\Sigma_{\rm cen-DM}/\Sigma_{\rm BCG+ICL})_i$, $\sigma$ is the intrinsic vertical scatter (along $y$-axis) of the model, and $N$ is the total number of points. Here we have assumed that the uncertainties in $x_i$ and $y_i$ are zero. This yields $(\alpha,\beta,\sigma)=(1.04,8.57,0.15)$ for $\Delta = 200$, and $(\alpha,\beta,\sigma)=(1.09,8.43,0.13)$ for $\Delta = 500$. The relations along with their 1-$\sigma$ scatters are shown using the solid black line and the dashed lines in Fig.~\ref{scaling}, respectively. 

From Fig.~\ref{scaling} we see a bending of the $\Sigma_{\rm cen-DM}/\Sigma_{\rm BCG+ICL}$ relationship at the outermost radii, with a stronger downturn in the case where we go beyond the $R_{500}$ and close to the $R_{200}$, eventually becoming nearly constant (more explicit in linear scale). This likely reflects the fact that the slopes of ICL and DM profiles become similar at higher radii \citep[e.g.][]{Pillepich2018}. This deviation from linearity could be reflected in the predictions made using the actual relation vs. the linear fits. We will explore this in the next subsections.

At this point, we would like to mention that our scaling relations may not be directly applicable to clusters because there are indications of systematic differences between observed $\Sigma_{\rm BCG+ICL}$ profiles and those in TNG300. This has been demonstrated for the \citet{Kluge2020} sample by \citet{Montenegro-Taborda2025b},
who found the observed SB profiles to be fainter by $\lesssim 0.9\,g'\,{\rm mag}\,{\rm arcsec}^{-2}$, with the exact offset depending on the radius. One would therefore need to correct for such differences before utilising the scaling relations. We will come back to this when we provide our recommendations for practical applications in Section~\ref{recom}.

In addition, we note that there are similar scaling relations between ICL and total matter (or DM) content based on other simulations reported in the literature. \citet{Alonso2020} used the \textsc{\large c-eagle} simulation \citep{Barnes2017} to derive a relation between the projected stellar mass density and projected total mass density for clusters with masses beyond $10^{14}\,{\rm M}_\odot$, and
found $\alpha=1.085$. \citet{Reina2023} presented a similar relation for haloes in the \textsc{\large e-mosaics} project \citep{Pfeffer2018} between ICL surface brightness and DM surface density, and reported
$\alpha=1.19$ for $M_{200}\gtrsim 10^{11}\,{\rm M}_\odot$ (the highest mass bin). More recently, \citet{Contreras2024} reported a slope of $\alpha=1.23$ for the relation between three-dimensional ICL and DM densities for clusters with masses $>8\times 10^{14}\,{\rm M}_\odot$ in the \textsc{\large the three hundred} simulation. 

The results by \citet{Alonso2020} and \citet{Reina2023} are for the $R_{200}$-normalised relations, and both show $\alpha$'s higher than ours (see the bottom-right corner of the panels in Fig.~\ref{scaling}). Likewise, the profiles in \citet{Contreras2024} were normalised by $R_{500}$ and also suggest $\alpha>1.03$. One possible reason for the lower slopes in our relations is the inclusion of BCG, which would reduce the multiplicative factor required to recover the DM content in the inner regions, resulting in a shallower slope for the relation. However, note that our relations are not directly comparable with these studies due to a variety of reasons: \citet{Alonso2020} showed the relation with stellar mass density instead of surface brightness; the sample in \citet{Reina2023} is significantly contaminated by haloes less massive than clusters; and the relation in \citet{Contreras2024} is based on three-dimensional stellar quantities. Furthermore, while \citet{Reina2023} utilised mock observations, they did not account for PSF convolution and noise. Hence, novel analyses with methodologies consistent across simulations are required to properly ascertain whether the slopes are indeed disparate.

\subsection{Recovered DM surface density profiles}
\begin{figure}
    \centering
    \includegraphics[width=1\columnwidth]{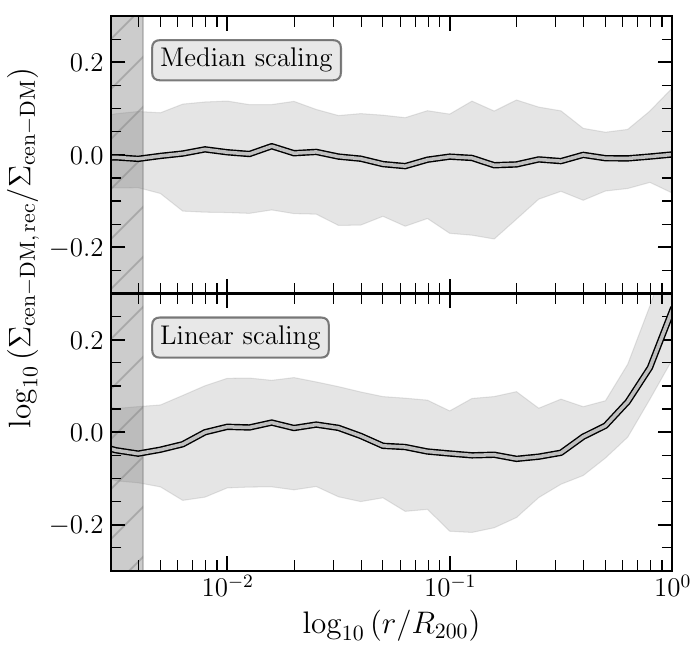}
    \caption{Comparison of $\Sigma_{\rm cen-DM}$ profiles recovered from the median 
    $\Sigma_{\rm cen-DM}$--$\Sigma_{\rm BCG+ICL}$ scaling relation for $R_{\Delta}=R_{200}$ (top
    panel) and the linear fit to this relation (bottom panel). Each panel shows the
    ratios of the recovered $\Sigma_{\rm cen-DM}$'s to the true values, where the solid 
    curve shows the median ratios and the shaded region shows the 16th to 84th percentiles. The
    vertical shaded region in the left spans the radii where the profiles are rendered unreliable
    due to two-body scattering between collisionless elements. The profiles recovered via
    the median scaling generally match the true ones, whereas those recovered with
    the linear scaling show systematic deviations, particularly at $r\gtrsim 0.5R_{200}$.}
    \label{comparerectotruth}
\end{figure}

Now that we have the scaling relations, we can use them to derive the DM halo profiles of the central
subhaloes ($\Sigma_{\rm cen-DM}$) from the ``observed'' BCG+ICL profiles. Here, we test how well are we able to recover the profiles for our clusters using the median scaling relations and their linear fits in Fig.~\ref{scaling}.
This is elucidated in Fig.~\ref{comparerectotruth}, where we show the results for $\Delta=200$. The top panel shows the ratio of the profiles recovered through the median scaling to the true profiles, and the bottom panel shows similar ratios for profiles recovered via the linear scaling. The solid curves
are the median ratios, and the shaded regions show their 16th to 84th percentiles.

The median ratios in the top panel are nearly constant at one across the whole dynamic range of $r/R_{200}$. The
ratios also exhibit a scatter of $\lesssim 0.15$~dex, as expected based on the linear fit. This means that, on average, the median scaling recovers the profiles well and devoid of any bias. The profiles obtained via the linear scaling, however, generally underpredict $\Sigma_{\rm cen-DM}$ at $r/R_{200}\lesssim 0.01$ and $0.03\lesssim r/R_{200}\lesssim 0.5$, and increasingly overpredict at higher radii for $r/R_{200}\gtrsim 0.5$. These systematics represent the offset of the linear fit with respect to the median scaling at different radii in Fig.~\ref{scaling}. The strongest deviation is the upturn at higher radii, which is caused by the fact that the $\Sigma_{\rm cen-DM}$--$\Sigma_{\rm BCG+ICL}$ relation essentially flattens in this range and the linear approximation is no longer representative.
This has implications for the DM concentrations and masses inferred for the central subhaloes, as we show next.

\subsection{Halo characteristics from the recovered DM profiles}\label{dmrec}
Our $\Sigma_{\rm cen-DM}$--$\Sigma_{\rm BCG+ICL}$ scaling relations can potentially be applied to real clusters to derive their central subhaloes' characteristics. This demands one to be mindful of the systematics that may impact the accuracy of this indirect inference. Considering this, we examine the utility of our relations by fitting the recovered $\Sigma_{\rm cen-DM}$ profiles to a halo model, and comparing the DM halo properties thus derived against the `true' values. 

We find that the three-dimensional DM profiles for the central subhaloes are, in general, well described by the Navarro-Frenk-White (NFW; \citealt{Navarro1996,Navarro1997}) formalism. Therefore, we opt for this model to estimate the halo concentration ($c_\Delta$) and mass ($M_\Delta$) from the surface density profiles. Specifically,
we fit the recovered $\Sigma_{\rm cen-DM}$ profiles to the \textit{projected} NFW model \citep{Lokas2001}:
\begin{equation}
    \Sigma_{\rm DM-NFW}(r) = \frac{{c_\Delta}^2\,g(c_\Delta)}{2\pi}\,\frac{M_\Delta}{{R_\Delta}^2}\,
    \frac{1 - |{c_\Delta}^2{\widetilde r}^2 - 1|^{-1/2}  A^{-1} [1/(c_\Delta\widetilde
    r)]}{({c_\Delta}^2{\widetilde r}^2 -1)^2}\,,    
\end{equation}
where
\begin{equation}
g(c) = \frac{1}{\ln (1+c_\Delta) - c_\Delta/(1+c_\Delta)}\,,
\end{equation}
\begin{equation} 
{\widetilde r} = r/R_\Delta\,,    
\end{equation}
and
\begin{equation}
    A^{-1} (x) = \left\{
    \begin{array}{ll}
    \cos^{-1} (x) & \mbox{if $r > R_\Delta/c_\Delta$} \\
    \cosh^{-1} (x)  & \mbox{if $r < R_\Delta/c_\Delta$ .}
    \end{array}
    \right.    
\end{equation}
Note that we only consider $r\gtrsim 0.004R_{\rm 200}$ when $\Delta=200$, and $r\gtrsim 0.007R_{\rm 500}$ when $\Delta=500$. The best-fit model is obtained through a Bayesian approach. Since the surface densities are based on discretised mass elements, we minimise the negative Poisson log-likelihood, that is, the $C$-stat function\footnote{Though this likelihood function is expected to asymptote to $\chi^2$ (Gaussian likelihood) for high number counts, such an approximation can lead to biased results \citep[e.g.][]{Humphrey2009}.}\citep{Kaastra2017}:
\begin{equation}\label{cstat}
C = 2 \sum\limits_{i=1}^{N} M_i - D_i + D_i \ln (D_i/M_i),
\end{equation} 
where $D_i=\log_{10}\Sigma_{{\rm cen-DM},i}$ and $M_i=\log_{10}\Sigma_{{\rm DM-NFW},i}$ is the datum and the model's value in $i$th bin, respectively, and $N$ is the total number of radial bins used to sample the profile. We use logarithmic surface densities in the convergence statistic, as it enables a more efficient exploration of the parameter space and better constraints on the quantities.

To determine the `true' DM parameters for the central subhaloes, we compute their spherically-averaged DM density profiles ($\rho_{\rm cen-DM}$) and fit them to NFW. For the latter, we fix the mass to the central subhalo's DM mass enclosed within $R_\Delta$, and minimise the statistic in equation~(\ref{cstat}) taking $D_i=\log_{10}\rho_{{\rm cen-DM},i}$ and $M_i=\log_{10}\rho_{{\rm DM-NFW},i}$. The usage of $\rho_{\rm cen-DM}$ instead of $\Sigma_{\rm cen-DM}$ is to avoid the uncertainties introduced by projection effects (see Appendix~\ref{2dvs3d}), and because the ultimate goal of the approach is to infer from observations the concentration parameter that is predicted from the N-body cosmological simulations.

\begin{figure*}
    \centering
    \includegraphics[width=1.8\columnwidth]{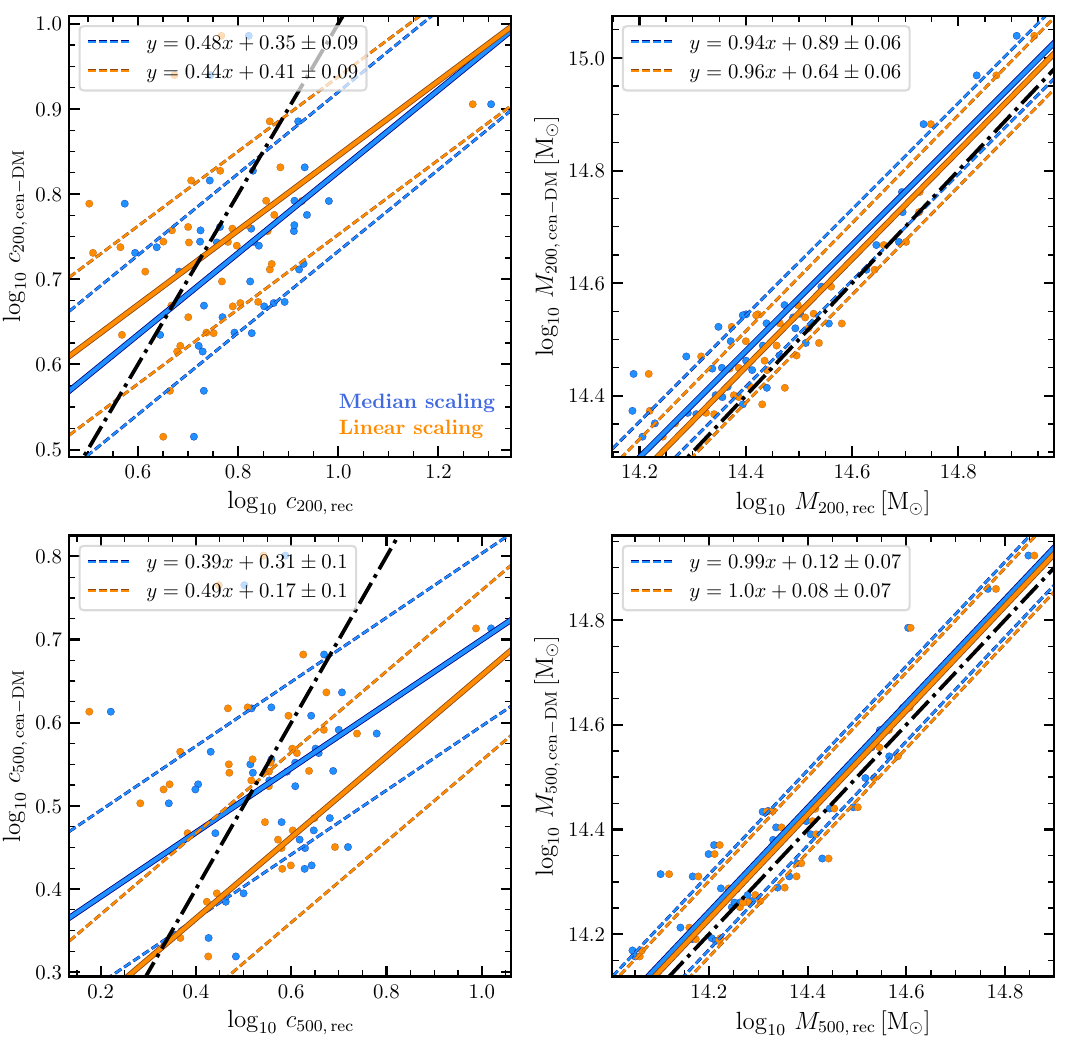}
    \caption{Comparisons of DM characteristics derived from the $\Sigma_{\rm cen-DM}$ profiles reconstructed from the BCG+ICL profiles (based on the masked optical images) against those derived from the 
    $\rho_{\rm cen-DM}$ profiles. All the DM properties correspond to the best-fit NFW profiles (Section~\ref{dmrec}), where the top row shows the results for the concentrations and halo masses assuming $R_{\Delta}=R_{200}$, and the bottom
    row shows similar results for $R_{\Delta}=R_{500}$. Each panel shows the values for 
    two approaches of reconstructing $\Sigma_{\rm cen-DM}$ from $\Sigma_{\rm BCG+ICL}$: a) using
    the raw median relation in Fig.~\ref{scaling} (shown in blue), and b) using the linear fit to the data in equation~(\ref{fit}) (orange). The solid orange/blue line corresponds to the best-fit for the
    data with the same colour, and the corresponding dashed lines encapsulate the 1-$\sigma$ scatter about the line. The black dash-dotted lines show the 1:1 relations.}
    \label{dmvsrecfits}
\end{figure*}

The DM concentrations and masses thus derived are compared in Fig.~\ref{dmvsrecfits}, where the top and bottom rows show the results for $\Delta=200$ and $\Delta=500$, and the left and right columns correspond to concentrations and masses, respectively. In each panel, the true central subhalo's parameters are plotted
on the vertical axis. The blue points show the DM properties indicated by the $\Sigma_{\rm cen-DM}$ profiles reconstructed using the median scaling relation (yellow curve; Fig.~\ref{scaling}), and the orange points are derived from the linear fit to the scaling relation [equation~(\ref{fit})]. In an ideal case, the points would lie in a 1:1 relation (black dash-dotted lines). We obtain linear fits to the orange and blue points, separately, which are shown as solid lines with the same colour correspondence. The dashed lines span the $1$-$\sigma$ regions. The coefficients and the scatter can be gleaned from the equations displayed in the top-left corner of each panel.

The top-left panel plots the $c_{200}$'s based on the $\rho_{\rm cen-DM}$ profiles ($c_{200,\,{\rm cen-dm}}$) against those from the recovered $\Sigma_{\rm cen-DM}$ profiles ($c_{200,\,{\rm rec}}$). The best-fit $c_{200,\,{\rm cen-dm}}$--$c_{200,\,{\rm rec}}$ relations clearly show that the predicted concentrations do not map directly
on to the true values. This discrepancy mostly stems from the scatter in Fig.~\ref{scaling} (and Fig.~\ref{comparerectotruth}), which in turn reflects the diversity in the phase-space distribution of ICL relative to that of DM. The varied formation times of the clusters is expected to play a role here. As a subhalo is accreted by a cluster, its DM is typically stripped earlier than stars, as 
the former tends to be more extended and less tightly bound gravitationally. 
Under this scenario, recently-assembled clusters have less time for the ICL to form or evolve. However, we find that this does not account for most of the scatter, which is rather stochastic in its origin (shown later in Section~\ref{scatter}). There is some additional contribution from projection effects (see Appendix~\ref{2dvs3d}).Nevertheless, as far as the recovery is concerned, the scatter indicates that it should be possible to derive the true values from the recovered ones within $\approx 0.09$~dex (i.e. $\approx 23$ per cent) uncertainty using the fitting relations that describe the
$c_{200,\,{\rm cen-dm}}$--$c_{200,\,{\rm rec}}$ relationships. 

The $M_{200,\,\rm cen-DM}$ predictions are shown in the top-right panel. The recovered values tend to lie above the 1:1 relation, that is, they are typically underpredicted. Note 
that the best-fit lines denote a scatter ($\approx 0.06$ dex) that is smaller than that for the concentrations. Thus, the $\Sigma_{\rm cen-DM}$--$\Sigma_{\rm BCG+ICL}$ scaling allows us to recover the central subhalo's DM mass with a significantly greater accuracy than its concentration. The minute systematic differences in the predictions can be easily accounted for using the best-fit relations to obtain the true masses.

The bottom row shows that the predictions obtained for $R_{\Delta}=R_{500}$ exhibit similar trends, with 
some differences. The slopes for the true vs recovered concentration relations from the median and the linear scalings exhibit a greater difference here. Also, the $M_{500,\,\rm cen-DM}$-$M_{500,\,\rm rec}$
relations are closer to 1:1. Overall, these
results indicate that, if we use the best-fit relations, the $c_{500,\,\rm cen}$'s can be recovered within an uncertainty of $\approx 0.1$~dex (or $\approx 26$ percent), and the $M_{500\,\rm cen}$'s can be derived within an uncertainty of $\approx 0.07$~dex (or $\approx 18$ percent).

It is worth noting here that the best-fit relationships for the values based on the true $\Sigma_{\rm cen-DM}$--$\Sigma_{\rm BCG+ICL}$ scaling and those from its linear approximation tend not to be drastically different, except for the concentrations based on $R_{\Delta}=R_{500}$. Furthermore, the predictions from the two kinds of scalings carry similar uncertainties. This is important with regard to application to real clusters, and suggests that one can utilise either one of these relations for this purpose. We will examine this in detail later in Section~\ref{global}.

Below, we summarise all the best-fit relations presented in Fig.~\ref{dmvsrecfits} for the convenience of the reader:
\begin{equation}\label{c200cenvsrec}
 \log_{10}c_{\rm 200,\,{\rm cen-DM}} = 0.48\log_{10} c_{\rm 200,\,rec-med} + 0.35 \pm 0.09,
\end{equation}
\begin{equation}\label{m200cenvsrec}
\log_{10}M_{\rm 200,\,{\rm cen-DM}} = 0.94\log_{10}M_{\rm 200,\,rec-med} + 0.89 \pm 0.06, 
\end{equation}
\begin{equation}\label{c500cenvsrec}
\log_{10}c_{\rm 500,\,{\rm cen-DM}} = 0.39\log_{10} c_{\rm 500,\,rec-med} + 0.31 \pm 0.1,
\end{equation}
\begin{equation}\label{m500cenvsrec}
\log_{10}M_{\rm 500,\,{\rm cen-DM}} = 0.99\log_{10}M_{\rm 500,\,rec-med} + 0.12 \pm 0.07,    
\end{equation}
\begin{equation}\label{c200cenvsrecl}
 \log_{10}c_{\rm 200,\,{\rm cen-DM}} = 0.44\log_{10} c_{\rm 200,\,rec-lin} + 0.41 \pm 0.09,
\end{equation}
\begin{equation}\label{m200cenvsrecl}
\log_{10}M_{\rm 200,\,{\rm cen-DM}} = 0.96\log_{10}M_{\rm 200,\,rec-lin} + 0.64 \pm 0.06, 
\end{equation}
\begin{equation}\label{c500cenvsrecl}
\log_{10}c_{\rm 500,\,{\rm cen-DM}} = 0.49\log_{10} c_{\rm 500,\,rec-lin} + 0.17 \pm 0.1,
\end{equation}
\begin{equation}\label{m500cenvsrecl}
\log_{10}M_{\rm 500,\,{\rm cen-DM}} = \log_{10}M_{\rm 500,\,rec-lin} + 0.08 \pm 0.07,    
\end{equation}
where `${\rm rec-med}$' in the subscript implies the usage of the median scaling relation, and `${\rm rec-lin}$' implies that the linear approximation is employed instead. 

\subsection{Scatter in the $\Sigma_{\rm cen-DM}$--$\Sigma_{\rm BCG+ICL}$ scaling relations}\label{scatter}
The uncertainty in the estimated quantities, for the most part, originates from the scatter in the scaling relations shown in Fig.~\ref{scaling}. This warrants a thorough exploration to identify the quantities that modulate this scatter. It would not only unravel the underlying physics that gives rise to the scatter, but can also aid in improving our predictions for halo characteristics -- the latter being contingent on the strength of correlations between the modulating factors and the scatter.

Clusters typically assemble by merging with other groups, bringing in additional satellites. Over time, both the stars and the DM in satellites are removed via tidal forces and/or accreted by the BCG.
Thus, the scatter in $\Sigma_{\rm cen-DM}/\Sigma_{\rm BCG+ICL}$ essentially reflects the diversity in the \textit{relative} growth of DM mass with respect to stellar mass in the central. 
One can therefore control for the scatter using a quantity that captures this difference in growth histories of DM and stars to an appreciable degree.

Note that DM is generally stripped with a greater efficiency than stars due to the former's larger extent and lower compactness \citep{Smith2016,Engler2021,Montero2024}. This implies that as a satellite enters a cluster, it looses DM first and the stars are liberated later. Under this scenario, a recently-formed cluster -- exhibiting a higher fraction of mass in satellites -- is expected to have a higher DM-to-stellar mass ratio in the central. 
Therefore, the scatter in $\Sigma_{\rm cen-DM}/\Sigma_{\rm BCG+ICL}$ is expected to be, at least in part, due to the diversity in the assembly histories of our clusters. In addition, given the halo mass span of our sample, we expect higher $\Sigma_{\rm cen-DM}/\Sigma_{\rm BCG+ICL}$ for massive haloes because such haloes tend to have formed recently \citep[e.g.][]{Li2008,McBride2009}, and also exhibit lower integrated star formation efficiencies \citep{Behroozi2013,Rodriguez2017,Kravtsov2018}.

Since stars in satellites are transferred to BCG+ICL as the cluster evolves, the fraction of total stellar mass of the cluster trapped in BCG+ICL is a useful observational proxy for its formation time \citep{Chun2023,Yoo2024,Montenegro-Taborda2025}. Also, the satellites lose their orbital angular momentum due to dynamical friction, whose strength increases with the satellite's mass. Hence, massive satellites are expected to reach the cluster centre earlier and get devoured by the BCG. This implies that one can assess a cluster's evolutionary state using the ratio between the $N$-th most massive satellite galaxy's mass and the BCG's mass \citep[e.g.][]{Deason2013,Kimmig2025,Montenegro-Taborda2025}. A photometric alternative is the difference in magnitudes, or the magnitude gap \citep{Dariush2010,Vitorelli2018,Golden-Marx2025}: $\Delta m_{1N}$, where $N$ refers to the $N$-th most luminous satellite.

\begin{figure*}
    \centering
    \includegraphics[width=1.55\columnwidth]{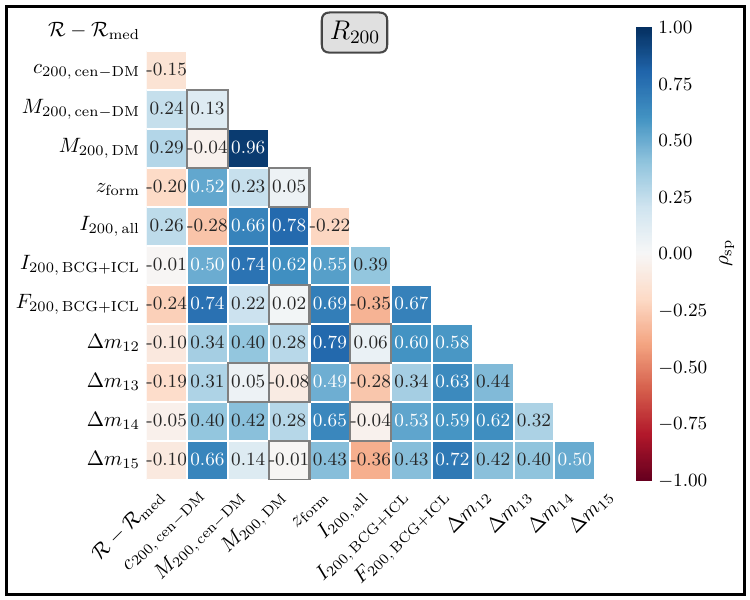}
    \includegraphics[width=1.55\columnwidth]{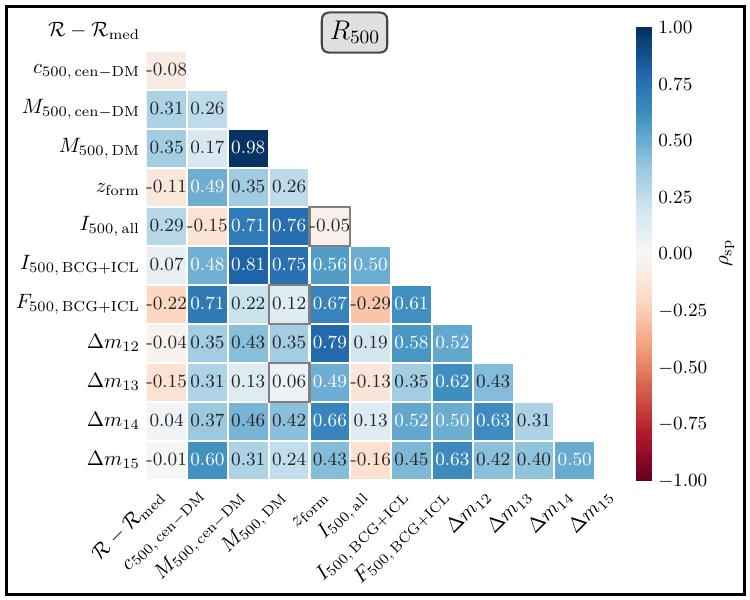}    
    \caption{Correlation matrices displaying the Spearman rank correlations of various
    halo and photometric quantities with the offset ($\mathcal{R}-\mathcal{R}_{\rm med}$; see the text) from one of the two median $\Sigma_{\rm cen-DM}$--$\Sigma_{\rm BCG+ICL}$ scaling relations in Fig.~\ref{scaling}, along with correlations among the quantities themselves. The top matrix shows the results for profiles extending out to $R_{200}$ (corresponding to the left panel in Fig.~\ref{scaling}), and the bottom matrix shows them for $R_{500}$. Each cell in a given matrix shows
    the correlation coefficient ($\rho_{\rm sp}$) for the associated pair of quantities, and corresponds to the colour indicated by the colour bar on the right. Trends with $<3$-$\sigma$
    significance are marked with grey boundaries around the respective cells. The offset is significantly correlated with
    the magnitude gaps, but at weak strengths ($\rho_{\rm sp} < 0.2$). The central subhalo's DM concentration correlates strongly ($\rho_{\rm sp}\gtrsim 0.68$) with the BCG+ICL fraction, $F_{\rm \Delta,\,BCG+ICL}$ (i.e. the fraction of total luminosity contributed by BCG+ICL). The central subhalo's DM mass exhibits the strongest trend (at $\rho_{\rm sp}\gtrsim 0.74$) with the BCG+ICL flux, $I_{\rm \Delta,\,{\rm BCG+ICL}}$.
    }
    \label{scatcor}
\end{figure*}

We examine how 11 DM/photometric quantities vary with offset from the \textit{median} scaling relation (yellow curves in Fig.~\ref{scaling}), and among themselves. This is carried out for both $\Delta=200$ and $\Delta=500$, separately. The quantities are described as
follows:
\begin{enumerate}
    \item $c_{\Delta,\,\rm cen-DM}$: DM concentration of the central subhalo based on the best NFW fit to the $\rho_{\rm cen-DM}$ profile extending out to $R_\Delta$ (same as in Fig.~\ref{dmvsrecfits}),
    \item $M_{\Delta,\,\rm cen-DM}$: DM mass of the central subhalo enclosed within $R_{\Delta}$,
    \item $M_{\Delta,\,\rm DM}$: Total DM mass of the cluster within $R_{\Delta}$,
    \item $z_{\rm form}$: Highest redshift (or the earliest time in its history) when the cluster accumulated more than 50 per cent of its present \textit{total} DM mass,
    \item $I_{\rm \Delta,\,all}$: Integrated flux of the unmasked image within the circular aperture of radius $R_\Delta$,
    \item $I_{\rm \Delta,\,BCG+ICL}$: Integrated flux of the masked image within the circular aperture of radius $R_\Delta$,
    \item $F_{\rm \Delta,\,BCG+ICL}$: BCG+ICL fraction within the circular aperture of radius $R_\Delta$, that is, $I_{\rm \Delta,\,BCG+ICL}/I_{\rm \Delta,\,all}$,  
    \item $\Delta m_{12}$: Magnitude gap or difference between the $g$-band magnitudes of the BCG and the most massive satellite\footnote{These magnitudes are taken from the TNG database and
    were computed for bound stars within 30 kpc spherical aperture, incorporating dust attenuation effects \citep{Nelson2018}.},
    \item $\Delta m_{13}$: Magnitude gap between the BCG and the second most massive satellite,
    \item $\Delta m_{14}$: Magnitude gap between the BCG and the third most massive satellite, and
    \item $\Delta m_{15}$: Magnitude gap between the BCG and the fourth most massive satellite.
\end{enumerate}

The offset from the median scaling relation at a given $r/R_{\Delta}$ is parameterised as $\mathcal{R}-\mathcal{R}_{\rm med}$, where $\mathcal{R}=\log_{10}\,(\Sigma_{\rm cen-DM}/\Sigma_{\rm BCG+ICL})$ for a cluster at a given $r/R_{\Delta}$, and $\mathcal{R}_{\rm med}$ is the expected ratio from the scaling. The strength and significance of trends are quantified through the Spearman rank correlation test, which is agnostic to the order of the correlation and accounts for non-linear monotonicities.

Note that the magnitude gaps opted for this analysis have been computed by considering the mass ranks for
\textit{all} satellites in the cluster, and are therefore agnostic to the $\Delta$ used to demarcate the cluster boundary. This is done so because, while the ICL may not always present sufficient signal-to-noise ratios at large radii in a realistic scenario, massive satellites are nonetheless detectable at those distances.

The results are shown using the correlation matrices in Fig.~\ref{scatcor}, with the top and
bottom matrices corresponding to $\Delta=200$ and $\Delta=500$, respectively. Each cell is coloured according to the Spearman rank correlation coefficient ($\rho_{\rm sp}$; mentioned within the cell) for the associated pair of quantities. The cell's boundary is shown in grey if the trend does not exhibit a $>3$-$\sigma$ significance (or $p$-value $<0.003$). We first focus on the trends based on $R_{\Delta}=R_{200}$.

The first column in the top matrix shows that the offset exhibits the strongest trends with $M_{200,\,\rm DM}$ and the total integrated (or bolometric) flux, $I_{200,\,\rm all}$. However, even for these properties, the correlation strength is low, with correlation coefficients $\rho_{\rm sp}=0.29$ and $\rho_{\rm sp}=0.26$, respectively. The above shows that the scatter in the $\Sigma_{\rm cen-DM}$--$\Sigma_{\rm BCG+ICL}$ scaling relation is not significantly segregated by any particular property.

The matrix also shows that $I_{200,\,{\rm all}}$ is strongly correlated with $M_{200,\,\rm DM}$ ($\rho_{\rm sp}=0.78$), as expected from the stellar-to-halo mass relation. As such, the offset's trend with $I_{200,\,{\rm all}}$ fundamentally represents the offset's dependence on 
the total DM mass. Likewise, we find a similar trend between the offset and the central's DM mass ($M_{200,\,{\rm cen-DM}}$; $\rho_{\rm sp}=0.24$), which is an excellent proxy for $M_{200}$. 

Next, in the decreasing order of correlation strength with the offset is the BCG+ICL fraction ($F_{200,\,{\rm BCG+ICL}}$; $\rho_{\rm sp}=-0.24$). Note that $F_{200,\,{\rm BCG+ICL}}$ and the formation time ($z_{\rm form}$) exhibit a strong positive correlation ($\rho_{\rm sp}=0.69$), that is, the BCG+ICL fraction is an indicator of the cluster's evolutionary state, such that later assembly corresponds to lower BCG+ICL fractions. This has also been demonstrated recently by \citet{Montenegro-Taborda2025} for $z=0$ in TNG300 clusters using the fractions based on stellar mass. For the reasons stated earlier, the magnitude gaps are also anticipated to encapsulate information about the assembly history. This is indeed evident in the positive correlations with $z_{\rm form}$ (fifth column, top matrix). In fact, we find that $\Delta m_{12}$ is the best predictor of $z_{\rm form}$ out of all the quantities for the massive clusters in our sample. The magnitude gaps also show negative trends with the offset, but at weaker strengths than $F_{200,\,{\rm BCG+ICL}}$.

Taken together, these trends with the offset suggest that clusters with higher masses, smaller BCG+ICL fractions, and smaller magnitude gaps have a weak preference for higher $\Sigma_{\rm cen-DM}$ for the 
same $\Sigma_{\rm BCG+ICL}$. This seems to align with the scenario where stellar stripping of an accreted satellite commences later than DM loss, and massive satellites reach the halo centre earlier, causing
the recently-assembled haloes to possess more DM mass in their central subhalo than stars. Since such haloes also tend to be more massive due to hierarchical assembly, massive haloes also exhibit higher $\Sigma_{\rm cen-DM}/\Sigma_{\rm BCG+ICL}$ ratios.

Interestingly, we also find that the concentration ($c_{200,\,\rm cen-DM}$) is strongly correlated with $F_{200,\,{\rm BCG+ICL}}$. \citet{Montenegro-Taborda2025} reported a similar correlation but for stellar quantities rather than fluxes. This is rather expected, considering that concentration represents
the mean density of the universe when the inner halo was assembled \citep{Wechsler2002,Zhao2003,Lu2006,Ludlow2013}. However, note that $c_{200,\,{\rm cen-DM}}$'s correlation with $F_{200,\,{\rm BCG+ICL}}$ ($\rho_{\rm sp}=0.74$) is considerably stronger than that with $z_{\rm form}$ ($\rho_{\rm sp}=0.52$). This highlights the additional role of concentration in the formation of BCG and ICL: a higher concentration implies a stronger tidal field near the halo centre, which is conducive for stripping at greater efficiencies \citep[e.g.][]{Contini2023,Martin2024} and directly contributes to the central's stellar growth. 

Similar results are obtained for $R_{\Delta}=R_{500}$ (bottom matrix; Fig.~\ref{scatcor}). This is demonstrated by the first column in the bottom matrix. Here, the strongest trends for the offset -- in decreasing order of strength -- are with the total DM mass ($M_{500,\,\rm DM}$), the central subhalo's DM mass ($M_{500\,{\rm cen-DM}}$), the bolometric flux ($I_{500,\,{\rm all}}$), the BCG+ICL fraction ($F_{500,\,{\rm BCG+ICL}}$, and the magnitude gap with the second most massive satellite galaxy ($\Delta m_{13}$). Likewise, $c_{500,\,\rm cen-DM}$ shows the strongest correlation with $F_{500,\,{\rm BCG+ICL}}$ and increases with $z_{\rm form}$.

Can we improve the inference of the DM distribution and DM halo parameters from the observed $\Sigma_{\rm BCG+ICL}$ profiles, as presented in previous subsections, by introducing more cluster (observable) properties? The answer to this question lies in the correlation strengths,
which convey that the aforementioned trends with the offset ($\mathcal{R}-\mathcal{R}_{\rm med}$) are rather weak, indicating that they are not useful for enhancing the predictive prowess of our relations. Most of the scatter, therefore, appears to be stochastic in nature, which also happens to be characteristic of ICL formation \citep{Cooper2015,Harris2017,Contini2023,Brown2024,Montenegro-Taborda2025}. Some of this is inherent to cluster assembly and unavoidable, because the exact radial mass distribution in the central subhalo depends on the initial positions and orbital parameters of the satellites, which are never identical for two clusters. Further complexity is introduced by the fact that satellites vary in their stellar-to-DM
mass and extent ratios. Moreover, the satellites accreted by a cluster via mergers with other groups have typically been pre-processed within those groups prior to their infall into the cluster \citep[e.g.][]{Donnari2021,Pallero2022,Manuwal2023,Park2023} -- implying that stellar stripping \textit{within the cluster} did not commence later than DM stripping for all the satellites, and a fraction of the ICL was formed outside the cluster.

Nevertheless, there are some promising insights gleaned from the correlation matrices that hint at the possibility of using alternate approaches to derive halo characteristics using photometry. For instance, we find that central
subhalo's DM mass is strongly correlated with the BCG+ICL flux. This could offer a straightforward way to estimate the central subhalo's DM mass with a reasonable degree of accuracy. Additionally, these results indicate that
we can use bolometric flux to estimate the cluster's \textit{total} DM mass, and BCG+ICL fraction
to derive the central subhalo's DM concentration. Next, we conduct a detailed exploration of the utility of these relationships.

\section{Dark matter properties derived through global photometry}\label{global}
\begin{figure*}
    \centering
    \includegraphics[width=2.05\columnwidth]{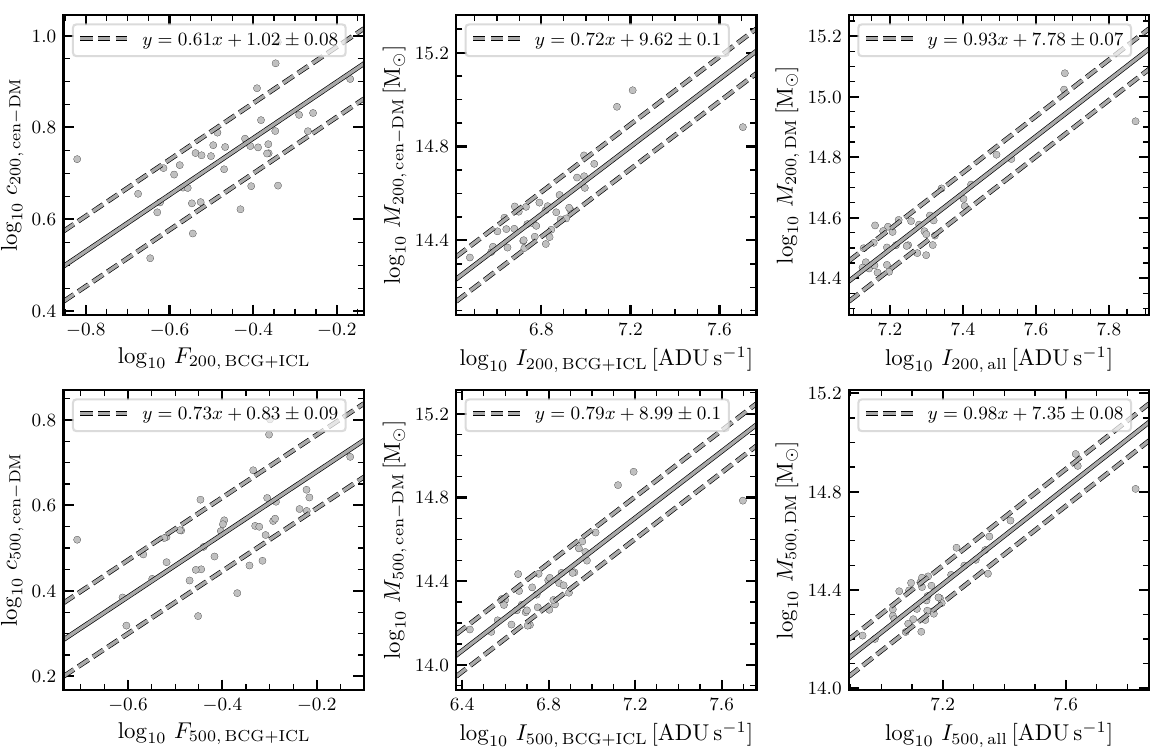}
    \caption{DM parameters for the clusters plotted against photometric quantities (defined in Section~\ref{scatter}) that exhibit the strongest correlations with those parameters (see Fig.~\ref{scatcor}). Panels in the top row show the results for when the halo boundary is taken as $R_{\Delta}=R_{200}$, and the bottom row similarly shows the
    results for $R_{\Delta}=R_{500}$. In each row, the left-most panel plots the relationship between the central subhalo's DM concentration and the BCG+ICL fraction of the cluster obtained through photometry, the middle panel plots the central subhalo's DM mass against the BCG+ICL flux,
    and the right-most panel plots the total DM mass against the bolometric flux. The solid line in each panel is the linear fit to the data, and the dashed lines span the $1$-$\sigma$ scatter about the fit. These relations offer an alternative route to infer the DM parameters of massive clusters via photometry.} 
    \label{dmvsphot}
\end{figure*}

The strong trends between DM halo parameters and photometric quantities suggested by Fig.~\ref{scatcor} 
can be used to estimate the former. We investigate this possibility in this section. First, we derive
the equations that best describe the relationships between the DM parameters and photometric measurements
(Fig.~\ref{dmvsphot}). Then, for the central subhalo, we compare the predictions from the relations
thus derived against those from equations~(\ref{c200cenvsrec})$-$(\ref{m500cenvsrecl}).

We begin by examining the relations for $R_{\Delta}=R_{200}$. The second
column of the top matrix in Fig.~\ref{scatcor} shows that $c_{200,\,{\rm cen-DM}}$ exhibits the strongest correlation with the light $F_{200,\,{\rm BCG+ICL}}$ fraction; a similar result has been reported in \citet{Montenegro-Taborda2025} for the corresponding 3D stellar mass fraction. We plot the logarithms of $c_{200,\,{\rm cen-DM}}$ and $F_{200,\,{\rm BCG+ICL}}$ in the top-left panel of Fig.~\ref{dmvsphot}, where it is evident that the two exhibit a tight linear relationship in this plane, shown by the best-fit line (solid grey). The corresponding equation in the top-left corner of the panel indicates that, by using this relation, the central subhalo's DM concentration can be determined within a 1-$\sigma$ uncertainty of $0.08$~dex. This 
scatter is smaller than the one in equations~(\ref{c200cenvsrec}) and (\ref{c200cenvsrecl}) by $0.01$~dex, indicating improvement in the predictive accuracy by just $\approx 2$ per cent.

Similarly, the third column of the correlation matrix indicates that $M_{200,\,{\rm cen-DM}}$ is
strongly correlated with $I_{200,\,\rm BCG+ICL}$. The relation, illustrated
in the top-middle panel of Fig.~\ref{dmvsphot} can aid in inferring $M_{200,\,{\rm cen-DM}}$ within $0.1$~dex. Note that this scatter is \textit{higher} than the one yielded by equations~(\ref{m200cenvsrec}) and (\ref{m200cenvsrecl}) by $\approx 10$ per cent.

The fourth column of the matrix implies that one can obtain $M_{200,\,{\rm DM}}$ of a cluster through $I_{200,\,\rm all}$. The best-fit relation in the top-right panel of Fig.~\ref{dmvsphot} presents a scatter of $0.07$~dex, that is, $I_{200,\,\rm all}$ can be used to estimate the total DM mass of the galaxy cluster with $\approx 18$ per cent uncertainty.

Similar relationships between these pairs of quantities are obtained for $R_{\Delta}=R_{500}$,
albeit with slightly ($0.01$~dex) larger scatters for the concentration and total DM mass. The $c_{500,\,{\rm cen-DM}}$--$F_{500,\,{\rm BCG+ICL}}$ relation shows a scatter of $0.09$~dex, which is again smaller than equations~\ref{c500cenvsrec} and \ref{c500cenvsrecl} by $0.01$~dex. The $M_{500,\,{\rm cen-DM}}$-$I_{500,\,\rm BCG+ICL}$ relation can be used to determine $M_{500,\,\rm cen-DM}$ within $0.1$~dex uncertainty, \textit{greater} than those in equations~(\ref{m500cenvsrec}) and (\ref{m500cenvsrecl}) by $0.03$~dex. Finally, $I_{500,\,\rm all}$ can be employed to infer the total DM mass within $0.08$~dex uncertainty.

Below, we lay out all the six relations in Fig.~\ref{dmvsphot}:
\begin{equation}\label{c200vsfbcgpicl}
 \log_{10}c_{\rm 200,\,\rm cen-DM} = 0.61\log_{10}F_{200,\,\rm BCG+ICL} + 1.02 \pm 0.08,
\end{equation}
\begin{equation}\label{m200vsibcgpicl}
\log_{10}M_{\rm 200,\,\rm cen-DM} = 0.72\log_{10}I_{200,\,\rm BCG+ICL}+ 9.62 \pm 0.1, 
\end{equation}
\begin{equation}\label{m200allvsiall}
\log_{10}M_{\rm 200,\,\rm DM} = 0.93\log_{10}I_{200,\,\rm all}+ 7.78 \pm 0.07, 
\end{equation}
\begin{equation}\label{c500vsfbcgpicl}
 \log_{10}c_{\rm 500,\,\rm cen-DM} = 0.73\log_{10}F_{500,\,\rm BCG+ICL} + 0.83 \pm 0.09,
\end{equation}
\begin{equation}\label{m500vsibcgpicl}
\log_{10}M_{\rm 500,\,\rm cen-DM} = 0.79\log_{10}I_{500,\,\rm BCG+ICL}+ 8.99 \pm 0.1, 
\end{equation}
\begin{equation}\label{m500allvsibcgpicl}
\log_{10}M_{\rm 500,\,\rm DM} = 0.98\log_{10}I_{500,\,\rm all}+ 7.35 \pm 0.08, 
\end{equation}

\begin{figure*}
    \centering
    \includegraphics[width=1.8\columnwidth]{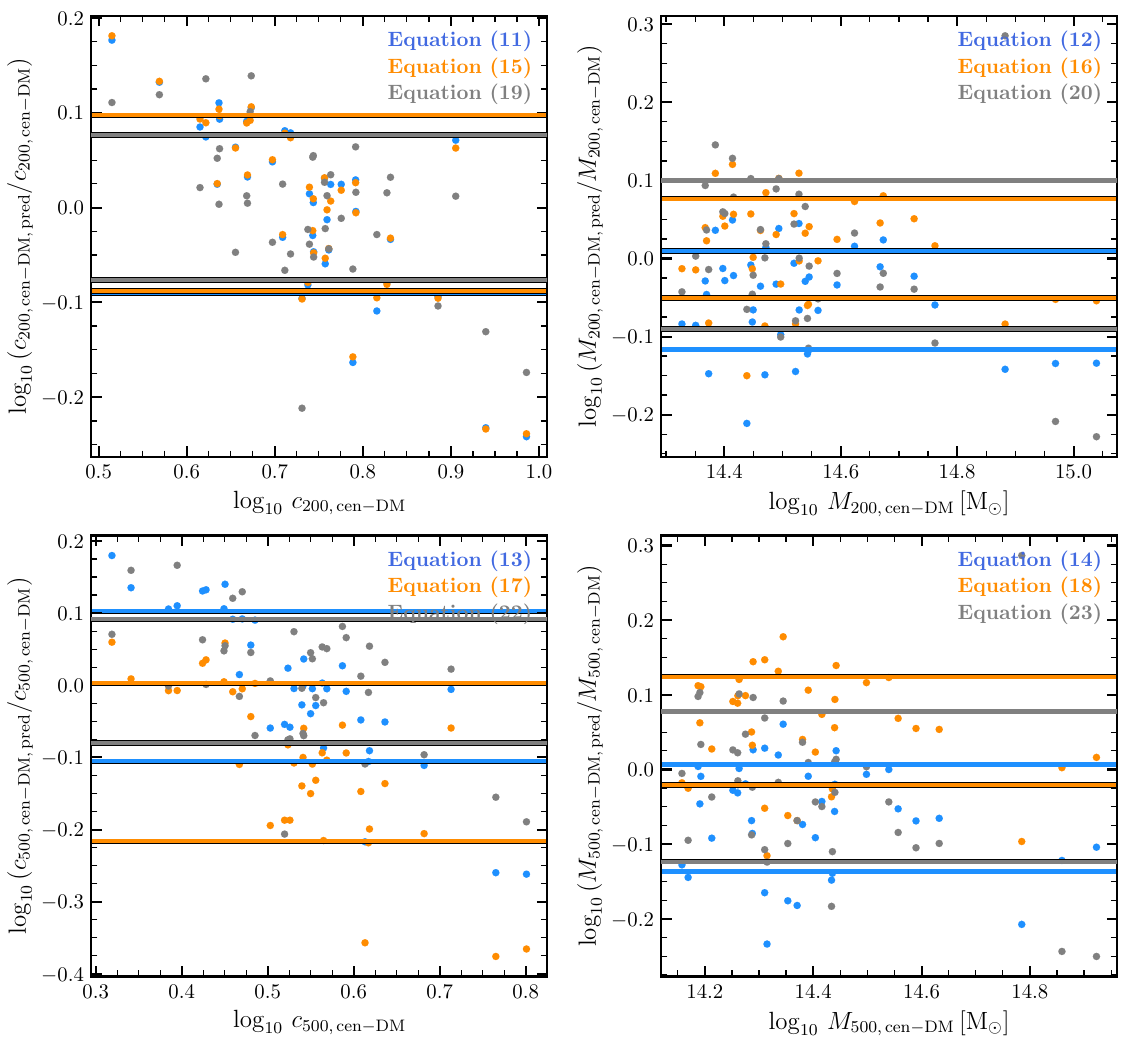}
    \caption{Comparison of predictions for DM characteristics for the central subhaloes in our TNG300 clusters derived from two approaches: $\Sigma_{\rm cen-DM}$--$\Sigma_{\rm BCG+ICL}$ scaling relation [equations~(\ref{c200cenvsrec})$-$(\ref{m500cenvsrecl})] and global photometry [equations~(\ref{c200vsfbcgpicl}), (\ref{m200vsibcgpicl}), (\ref{c500vsfbcgpicl}), and (\ref{m500vsibcgpicl})]. The values from the former method are either based on the true median
    scaling relation or the linear fit to the scaling, and are shown in blue and orange, respectively.
    The predictions from global photometric measurements appear in grey. The top and bottom rows correspond to parameters for $R_{\Delta}=R_{200}$ and $R_{\Delta}=R_{500}$, respectively. Each panel plots the ratio of predicted quantity to its true value against the true value, with those in the left column showing the concentrations, and the right column showing the masses. The horizontal solid lines demarcate the 1-$\sigma$ regions.}
    \label{comparepred}
\end{figure*}

As an additional exercise, we now directly apply the two methods to our clusters and compare the predictions against the true DM parameters for the central subhalo. The motivation here is to understand how are the results expected to vary in practice depending on the opted method. The comparisons are displayed in Fig.~\ref{comparepred}, where a given row corresponds to $\Delta=200$ or $\Delta=500$, and each panel shows the ratio of the predicted value for the quantity to its true value (derived from the central subhalo's $\Sigma_{\rm DM}$ profile). The values obtained using the median $\Sigma_{\rm cen-DM}$--$\Sigma_{\rm BCG+ICL}$ scaling [equations~(\ref{c200cenvsrec})$-$(\ref{m500cenvsrec})] are shown in blue, those
from the linear scaling [equations~(\ref{c200cenvsrecl})$-$(\ref{m500cenvsrecl})] are shown in orange, 
while those based on global photometry [equations~(\ref{c200vsfbcgpicl}), (\ref{m200vsibcgpicl}), and (\ref{c500vsfbcgpicl})] are displayed in grey. In each panel, the solid horizontal lines encompass
the 1-$\sigma$ regions.

First, we focus on the results for $R_{200}$ in the top row. The left panel
shows that the $c_{200,\,\rm cen-DM}$ predictions are, on average, close to the true values regardless of the chosen method (the mean ratios are $\approx 1$). The scatter is slightly greater for the scalings,
but the Levene tests show that this increase is not significant ($p$-values $\approx 0.25$). Thus, the methods can be considered equivalent for inferring the concentrations.

For the DM masses (top-right panel), we find that the scaling methods present slightly smaller scatters but the differences are insignificant (Levene test $p$-values $\approx 0.09$). The panel also shows that the values are typically biased low by $\approx 0.05$~dex for the median scaling (blue). This is also supported by the Welch's t-tests at a $\gtrsim 3$-$\sigma$ level. Such biases essentially indicate that the mapping between $M_{200,\,\rm cen-DM}$ and $M_{200,\,\rm rec}$ (top-left panel, Fig.~\ref{dmvsrecfits}) is not perfectly linear. The linear
scaling and the global photometry therefore seem more preferable for deriving $M_{200,\,\rm cen-DM}$'s.

Next, we examine the results for $R_{\Delta}=R_{500}$ in the bottom row of Fig.~\ref{comparepred}. For
$c_{500,\,\rm cen-DM}$, we find that the median scaling (blue) presents a scatter greater than that via equation~(\ref{c500vsfbcgpicl}) by $\approx 0.02$ dex, but this is not a significant difference ($p$-value for the Levene test is $\approx 0.4$). Also, the average prediction matches the true concentration for both. The linear scaling, however, underpredicts by $\approx 0.1$~dex, which is a strong discrepancy ($p$-values for the Welch's $t$-tests are $\sim 10^{-5}$). Therefore, both the median scaling and the global photometry methods are preferable over the linear scaling.

Finally, the bottom-right panel of Fig.~\ref{comparepred} shows that the $M_{500\,\rm cen-DM}$ predictions from the scaling relations exhibit biases of $\approx 0.05$~dex, whereas the ones derived from $I_{500\,\rm BCG+ICL}$ demonstrate a two-folds smaller bias of $\approx -0.025$~dex. However, Welch t-tests show that 
the difference between the mean ratios for the global photometry and the median scaling are not at a 3-$\sigma$ level. The two are also consistent in their scatters (the Levene test gives $p$-value $\approx 0.18$). Thus, we recommend avoiding the linear scaling for obtaining $M_{500,\,\rm cen-DM}$'s

Note that, though the relationships based on global photometry generally result in smaller biases in the predicted quantities, one should take caution while applying these to observed images that are seriously
contaminated by foreground objects (like Galactic stars). This is because such objects are masked during the analysis, and overmasking precludes accurate estimations of bolometric flux. The $\Sigma$ profiles 
are expected remain unaffected by such contaminations, because they capture the \textit{median} brightness within the annuli, and the masks tend to be spatially homogeneous (like those in \citealt{Montenegro-Taborda2025b}, for example). In such cases, using the $\Sigma_{\rm cen-DM}$--$\Sigma_{\rm BCG+ICL}$ scaling relations is rather preferable.

\section{Caveats \& recommendations}\label{caveats}
As stated in Section~\ref{sim}, \tng produces realistic clusters 
and the galaxies therein over a wide range of properties. This provides substantial support for the reliability of the simulation used in this work for our science case. There nonetheless remain certain numerical/physical effects that may limit the utility of our results for direct application to observations. For instance, changing the resolution of mass elements can have important consequences for the build up of stellar mass \citep{Pillepich2018,Pillepich2018b,Engler2021,Martin2024}. Similarly, the introduction of baryons 
into DM haloes can modify the profiles \citep{Lovell2018}. It is important to be mindful of the exact contribution from these effects for an informed implementation of our results. Motivated by this, we quantify the impact of such factors on the radial profiles of our clusters. We describe the resolution effects in Section~\ref{reseffect}, and the aftermath of introducing baryons into the simulation in Section~\ref{bareffect}. In addition, we provide our final recommendations for applications to real
clusters in Section~\ref{recom}.

There may also be some inaccuracies due to the limitations inherent to \textsc{\large subfind} \citep[see][]{Moreno2025}. A detailed exploration would require comparing results from different substructure-finding algorithms, which is beyond the scope of this work. We therefore defer this for future studies.

\subsection{Mass resolution}\label{reseffect}
We begin by investigating the impact of the resolution of mass elements. The simulation used in this work presents the finest resolution among all the TNG300 runs, which precludes us from comparing directly against
higher resolution versions of our clusters. We therefore utilise the runs for TNG100, namely, TNG100-1 and TNG100-2 \citep[see][]{Nelson2019}. The latter has the same resolution
as the simulation employed by us, but the former exhibits a 10 times better resolution. For this exercise,
we select the 14 most massive clusters above $M_{200}=10^{14}\,{\rm M}_\odot$ in these runs, and compute
profiles for the centrals therein. Specifically, we obtain three kinds of profiles: DM density ($\rho_{\rm cen-DM}$), DM surface density ($\Sigma_{\rm cen-DM}$), and stellar surface density ($\Sigma_{\rm cen-\star}$).
Then, we take the ratio of the profiles obtained for the haloes in higher resolution run to their counterparts
in the lower resolution run.  

\begin{figure}
    \centering
    \includegraphics[width=1\linewidth]{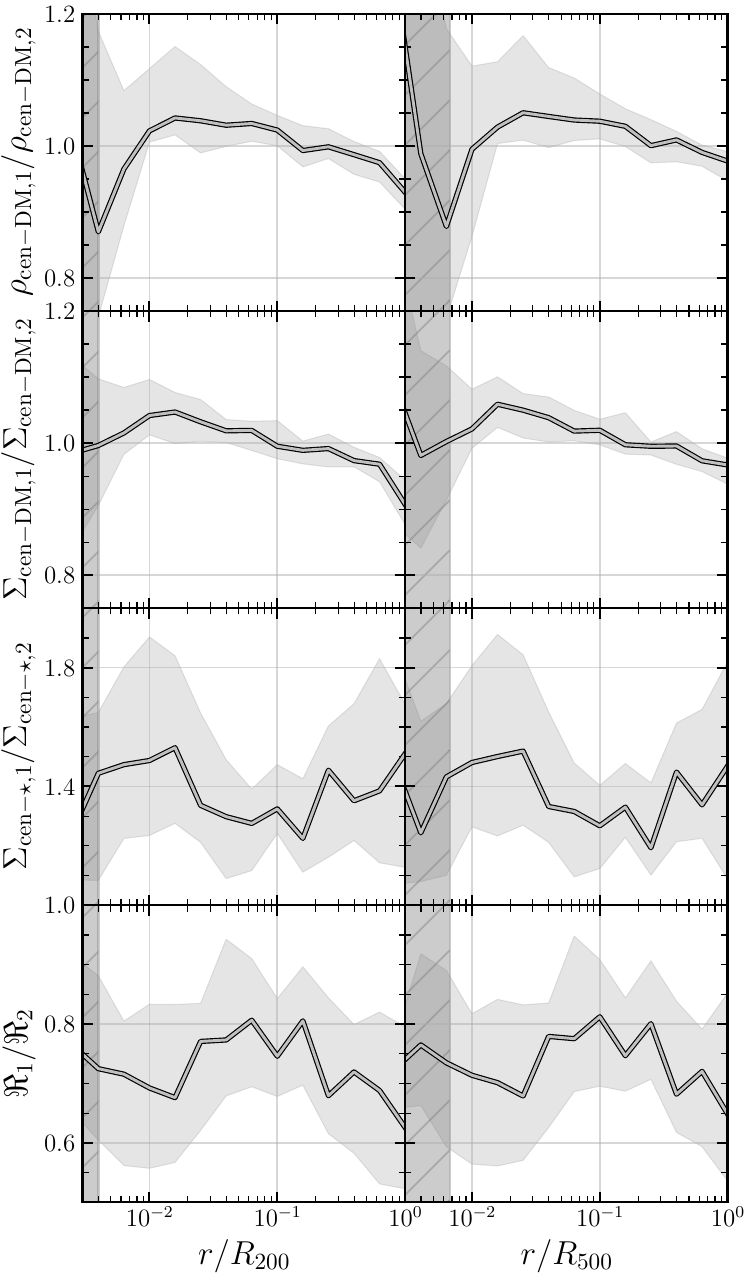}
    \caption{Impact of particle mass resolution on the stellar and DM profiles of central subhaloes
    within the most massive clusters ($M_{200}>10^{14}\,{\rm M}_\odot$) in TNG100. Each row corresponds
    to a specific type of profile, and shows the ratio of profile in the higher resolution run (TNG100-1) to that in the lower resolution run (TNG100-2; the same resolution as TNG300). Perusing from top to bottom, the panels show the ratios of DM density ($\rho_{\rm cen-DM}$), DM surface density ($\Sigma_{\rm cen-DM}$), stellar surface density ($\Sigma_{\rm cen-\star}$), and the scaling between DM and stellar surface densities ($\Re=\Sigma_{\rm cen-DM}/\Sigma_{\rm cen-\star}$). The curves are the median ratios and the shaded regions surrounding them span 16th to 84th percentiles. The vertical strips on the left denote the radii affected by numerical heating. The left and right columns show the results for $\Delta=200$ and $\Delta = 500$, respectively.}
    \label{res}
\end{figure}

The ratios are presented in Fig.~\ref{res}, where each row corresponds to certain kind of profile,
and the columns show the profiles with radii normalised against $R_{200}$ (left) or $R_{500}$ (right). From
top to bottom, the panels show the results for $\rho_{\rm cen-DM}$, $\Sigma_{\rm cen-DM}$, $\Sigma_{\rm cen-\star}$, and $\Re=\Sigma_{\rm cen-DM}/\Sigma_{\rm cen-\star}$. The curves show the medians and the surrounding
shaded regions encompass 16th to 84th percentiles. The vertical shaded strips on the left of each panel
demarcate the radii affected by spurious numerical heating (see Section~\ref{uniscale}), and we only consider
the trends beyond these radii.

The top-most row shows that the $\rho_{\rm cen-DM}$ profile from TNG100-1 generally deviates compared
to its TNG100-2 analogue by $\lesssim 5$~per cent, except at $r\lesssim 0.01R_\Delta$ where the differences
can reach $\approx 10$ per cent. More importantly, the nature of this deviation is radius-dependent
such that inner-most regions are typically denser, and outer regions are somewhat underdense at higher resolution. Similar results are seen for the $\Sigma_{\rm cen-DM}$ profiles in the panels below, with one key difference at $r/R_\Delta\lesssim 0.01$, where we now see an \textit{enhancement} 
by $\lesssim 5$ per cent. The associated rise in DM concentration at higher resolution is, therefore, expected to be even greater when derived from the $\Sigma_{\rm cen-DM}$ profiles.

Next, we shift our focus on the third row from the top, which shows the changes incurred in the $\Sigma_{\rm cen-\star}$ profiles. The panels demonstrate that the increase in resolution is
accompanied by an increase in $\Sigma_{\rm cen-\star}$ at all radii, and that this systematic varies -- on average -- between $\approx 20-50$ per cent. This is similar to the findings of \citet{Pillepich2018b}, who compared the stellar masses and cumulative stellar mass profiles of the haloes with $10^{13}\leq M_{200}/{\rm M}_\odot \leq 10^{14}$. They inferred a correction factor of $\approx 1.4$ for
the masses and showed that applying this to the $M_\star$'s of clusters with $M_{200}>10^{14}\rm M_\odot$ suffices to bring TNG300 and TNG100-1 in agreement. For the profiles, this is achieved by applying
radius-dependent correction factors. 

The rise in stellar content with resolution stems from the fact that the threshold density for star formation in \tng is fixed to the same value for all resolutions \citep[see appendix A in][]{Pillepich2018}. While this is reasonable for examining `strong convergence' \citep{Schaye2015}, it also inevitably leads to higher star formation rates because the simulation samples regions with greater densities than those in lower resolution runs. This generally exacerbates even further at higher resolutions, but there is appreciable convergence at the TNG100-1 resolution for the masses spanned by the clusters in this paper \citep{Engler2021}.

The results for $\Sigma_{\rm cen-\star}$ profiles can be considered directly translatable to BCG+ICL surface brightness profiles (barring the inaccuracies due to masking), as the mass-to-light ratio profiles for clusters are agnostic to numerical resolution. Taken together with the $\Sigma_{\rm cen-DM}$ profiles, this has important implications for the $\Sigma_{\rm cen-DM}/\Sigma_{\rm BCG+ICL}$ scaling. We examine this explicitly in the bottom-most row of Fig.~\ref{res}, where we show the changes in $\Re\equiv\Sigma_{\rm cen-DM}/\Sigma_{\rm cen-\star}$ scaling due to resolution. The median curves appear like inversions of those in the panels immediately above, indicating that the changes in the scaling mainly reflect those in the stellar profiles. This is also indicated by the values of the medians at $\approx 0.65-0.80$, and is indeed expected given that the changes imparted in $\Sigma_{\rm cen-DM}$ are relatively minor.

\begin{figure}
    \centering
    \includegraphics[width=1\linewidth]{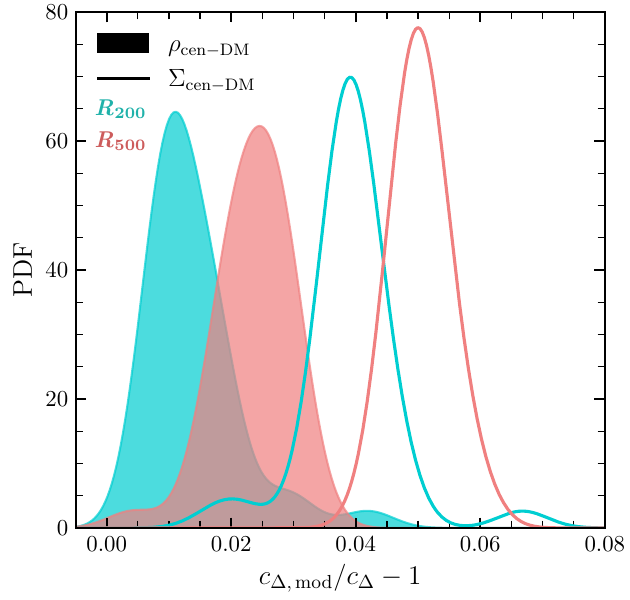}
    \caption{Proportional changes in the DM concentrations of our TNG300 clusters predicted due to resolution effects. The figure shows difference between the NFW concentrations obtained for the DM profiles of central subhaloes modified by the median factors in Fig.~\ref{res} ($c_{\Delta,\, \rm mod}$), and the concentrations suggested by the default profiles from the simulation, normalised by the latter. The filled and open histograms show the distributions for 
    $\rho_{\rm cen-DM}$ and $\Sigma_{\rm cen-DM}$, respectively. The values for the two choices of halo extent
    ($R_{\Delta}$) are differentiated using specific colours.}
    \label{resconc}
\end{figure}

We now extend our analysis to explicitly assess the degree of variation in the inferred concentrations that such resolution corrections are predicted to cause. For this, we apply the median corrections 
in the top rows of Fig.~\ref{res} to the $\rho_{\rm cen-DM}$ and $\Sigma_{\rm cen-DM}$ profiles of our TNG300 clusters, and deduce the concentrations of these modified profiles through the best-fit NFW models. In Fig.~\ref{resconc}, we show the probability distribution functions (PDFs) for the fractional difference between these newly inferred values
($c_{\Delta,\,\rm mod}$) and the original values. The turquoise and red histograms correspond to profiles extending upto $R_{200}$ and $R_{500}$, respectively. For each $R_\Delta$, there is a filled and an open histogram, showing the results for the values based on $\rho_{\rm cen-DM}$ and $\Sigma_{\rm cen-DM}$ profiles, respectively.

It is clear from these PDFs that increasing the resolution typically increases the concentration, as is indeed expected given the fact that the profiles exhibit higher densities at most of the inner radii, and lower densities in the outskirts (Fig.~\ref{res}). We would like to emphasise, however, that these changes are rather minor amounting to just $\lesssim 7$ per cent, meaning that the concentration measurements are
relatively robust against resolution effects. Interestingly, the PDFs also show that the magnitude of enhancement differs systematically depending on the adopted $R_\Delta$ and the type of DM profile ($\rho$ or $\Sigma$).

The fractional increase for the $\Sigma_{\rm cen-DM}$ profiles (open curves) ranges
between $\approx 2-7$ per cent, whereas that for $\rho_{\rm cen-DM}$ is $\lesssim 4$ per cent. 
The fundamental reason for this is that projected densities at smaller radii have greater contributions from radii larger than the probed scale. For a given profile type, the impact is stronger for the $c_{500}$ (red) than $c_{200}$ (turquoise), but this is a two-folds weaker effect (at $\approx 1.5$ per cent) than the difference between the profile types.

\subsection{Baryonic effects}\label{bareffect}
Next, we carry out a similar exploration to quantify the impact of baryons on the DM profiles. This
essentially deals with the changes in gravitational potential caused by the presence of baryons and the
associated hydrodynamical processes (like feedbacks). To test this, we first identify the counterparts
of our clusters in the dark-matter-only (DMO) variant of TNG300. We use the IDs of matched haloes
in the TNG data base that were derived via bijective cross-matching of particles \citep{Rodriguez-Gomez2015}. Once identified, we determine the DM density and surface density profiles of their central subhaloes.

\begin{figure}
    \centering
    \includegraphics[width=1\linewidth]{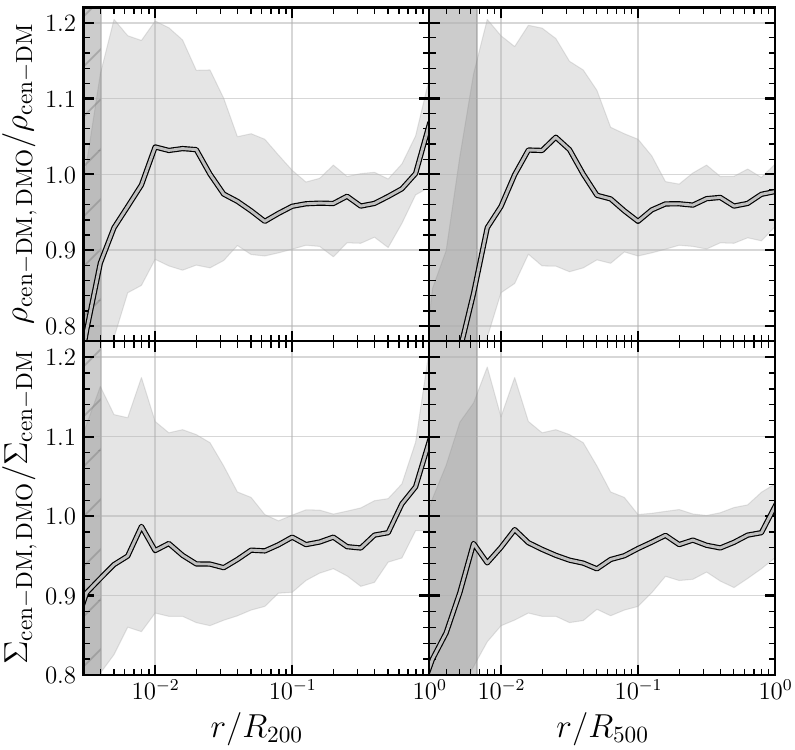}
    \caption{Changes in our DM profiles induced by the introduction of baryons. The top and bottom rows
    show the results for $\rho_{\rm cen-DM}$ and $\Sigma_{\rm cen-DM}$) profiles, respectively. Each panel
    shows the ratio of the profile obtained for the cluster's counterpart in the dark-matter-only (DMO) run
    of TNG300 to the profile in the hydrodynamical run. The curves are the median ratios and the shaded regions surrounding them span 16th to 84th percentiles. The vertical strips on the left denote the radii affected by numerical heating. The left and right columns show the results for $\Delta=200$ and $\Delta=500$, respectively.}
    \label{bar}
\end{figure}

Fig.~\ref{bar} shows the ratio of these profiles to those derived from the full physics (FP) run,
with the top row showing the results for $\rho_{\rm cen-DM}$ and the bottom row for $\Sigma$ profiles. The
curves and shaded regions have the same meaning as in Fig.~\ref{res}. The top
panels implies reduced $\rho_{\rm cen-DM}$ at most scales with a degree of suppression that 
reduces with normalised radius beyond $r\approx 0.1R_\Delta$. Similar results are seen for $\Sigma_{\rm cen-DM}$
profiles in the panels below, but without the enhancement at $0.01\lesssim r/R_\Delta\lesssim 0.02$ observed
for $\rho_{\rm cen-DM}$, and a weaker suppression at smaller radii. This difference between three dimensional and surface densities (like that in Fig.~\ref{res}) is again due to the substantial contribution from higher radii in projected densities measured at small scales.

\begin{figure}
    \centering
    \includegraphics[width=1\linewidth]{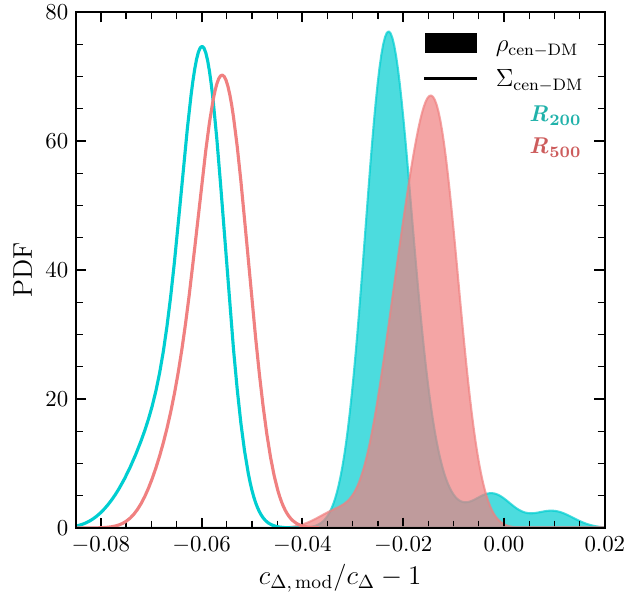}
    \caption{Impact of baryons on the DM concentrations of our clusters. Each histogram
    shows the probability distribution function for the relative change in the DM concentration for the central
    on switching from the hydrodynamical run to the DMO run. The filled and open histograms correspond to
    the values based on $\rho_{\rm cen-DM}$ and $\Sigma_{\rm cen-DM}$ profiles, respectively. The colours
    denote different $R_\Delta$ choices.}
    \label{barconc}
\end{figure}

We show the ramifications of these profile modifications for the concentration estimates in Fig.~\ref{barconc}. The concentrations used in the numerator are based on the best-fit NFW models to the DM profiles of the DMO analogues of our clusters. These fractional changes broadly indicate that removal
of baryons from the simulation leads to concentrations that are lower by $\lesssim 8$ per cent in most cases. This 
reflects the adiabatic contraction \citep{Gnedin2004,Duffy2010,Anbajagane2022b,Sorini2025} and ties back to the results presented in Fig.~\ref{bar}. The median curves show an overall rising 
density ratio with $r/R_\Delta$, meaning that the DMO profiles are shallower. This trend is exhibited by both $\rho$ and $\Sigma$ profiles, but the weaker deviations for the latter at small radii effectively cause shallower profiles, thereby resulting in $\approx 4$ per cent greater reduction in the halo concentration.

\subsection{Recommendations for applications to observed clusters}\label{recom}
Here, we describe our suggested approaches for leveraging the results in this paper to obtain DM distribution and parameters for real clusters. This has been informed based on our findings in the above sections, and some earlier works like \citet{Pillepich2018b} and \citet{Montenegro-Taborda2025b}.

It is important to note that the relations presented in this work implicitly require \textit{a priori} estimates for $R_{200}$ and/or $R_{500}$, which also implies measurements of total (baryons+DM) mass within these radii. Our results are for deducing the DM properties specifically, and for the clusters that 
are massive enough to lie in the range explored here.

For the $\Sigma_{\rm cen-DM}/\Sigma_{\rm BCG+ICL}$ radial scalings (Fig.~\ref{scaling}), we 
recommend using the median scalings. Note that centrals in TNG300 clusters are likely to be systematically
brighter/more-massive than observed ones \citep[e.g.][]{Pillepich2018,Pillepich2018b,Montenegro-Taborda2025b}. Hence, we suggest employing our median scalings only after applying the appropriate corrections to the observed $\Sigma_{\rm BCG+ICL}$ profiles, to account for the differences between simulated and observed SB profiles. (For WWFI-like observations, the reader can use the median $\Sigma_{\rm BCG+ICL}$ profile for our TNG300 clusters that we provide with this paper.) The corrected, observed BCG+ICL profiles and the scaling relations should then be used to derive the DM profiles, and the corresponding DM parameters should be obtained through the best-fit projected NFW models and equations~(\ref{c200cenvsrec})-(\ref{m500cenvsrec}). At last, one must add additional uncertainties on the inferred concentrations to account for baryonic and resolution effects in accordance with the systematics presented in Figs.~\ref{resconc} and \ref{barconc}. For the readers interested in the detailed $\Sigma_{\rm cen-DM}$ profile in addition to the DM parameters, 
we provide the median ratios shown in Figs~\ref{res} and~\ref{bar} for gauging uncertainties in the profile due to resolution effects and baryonification.

For the relationships with global properties in Fig.~\ref{dmvsphot}, the approach depends on the relationship in question. Considering that the \textit{total} stellar content is consistent with observations \citep{Pillepich2018b}, we believe the relationships between total DM mass and total integrated flux
(right-most column, Fig.~\ref{dmvsphot}) can be applied directly. However, since the BCG+ICL mass is
higher in the simulations, the relations with BCG+ICL flux and BCG+ICL fraction should be applied
after correcting for this offset. Furthermore, it is important to incorporate the uncertainties from resolution and baryonic effects while handling the concentrations obtained from $c_{\Delta,{\rm cen-DM}}$--$F_{\Delta,\,{\rm BCG+ICL}}$ relations.

\section{Summary and Conclusions}\label{conclude}
In this study, we have analysed the 40 most-massive galaxy clusters ($M_{200}\gtrsim 10^{14.5}\,{\rm M}_\odot$) 
at $z\approx 0.06$ from the TNG300 simulation from the \tng suite \citep{Weinberger2017,Pillepich2018},
aiming to assess the potential of deep optical imaging to infer the DM properties of galaxy
clusters. To this end, we generated mock optical observations for the $g'$ band of Wendelstein Wide-Field Imager (WWFI; \citealt{Kosyra2014}) onboard the 2-m Fraunhofer telescope at the Wendelstein Observatory,
taking the observations by \citet{Kluge2020} as our reference (Section~\ref{imagegen}). 

We isolated the emission from BCG and ICL by masking the satellites (Section~\ref{isolate}),
and used these masked images to compute the BCG+ICL surface brightness profiles for our clusters ($\Sigma_{\rm BCG+ICL}$; Section~\ref{analysis}). Similarly, we calculated the DM surface density profiles for the central subhalo ($\Sigma_{\rm cen-DM}$) derived from the projected DM mass map. These profiles were used to obtain the scaling relations between $\Sigma_{\rm cen-DM}$ and $\Sigma_{\rm BCG+ICL}$ profiles for two different choices of cluster extent ($R_{\Delta}=R_{200}$ and $R_{500}$; Fig.~\ref{scaling}). Then, we obtained the concentrations and masses from the $\Sigma_{\rm cen-DM}$ profiles reconstructed from these scaling relations and devised equations to convert these to the true values based on the three-dimensional DM density profiles [equations~(\ref{c200cenvsrec})$-$(\ref{m500cenvsrecl}); Section~\ref{dmrec}]. Additionally, we investigated potential sources of scatter in the $\Sigma_{\rm cen-DM}$--$\Sigma_{\rm BCG+ICL}$ scaling relations (Section~\ref{scatter}). Inspired from these results, we explored the prospect of recovering central subhalo's DM characteristics and total DM mass of clusters via the integrated flux measurements that exhibit the strongest correlations with these properties (Section~\ref{global}). 
Finally, we quantified the impact of resolution and baryonic effects on our profiles, 
and provided recommendations for applications to real clusters (Section~\ref{caveats}).

The salient points from this work are the following:
\begin{itemize}
    \item The $\Sigma_{\rm cen-DM}$ profile exhibits a quasi-linear scaling relation with the profile, whether the halo extent ($R_{\Delta}$) is considered to be $R_{200}$ or $R_{500}$ (Fig.~\ref{scaling}; Section~\ref{uniscale}). The best-fit linear relation 
    corresponds to $(\alpha,\beta,\sigma)=(1.05,8.57,0.15)$ for $\Delta = 200$, and $(\alpha,\beta,\sigma)=(1.09,8.43,0.13)$ for $\Delta = 500$, where $\alpha$ is the slope, $\beta$ is the intercept, and $\sigma$ is the vertical scatter. These scaling relations can be used to recover the central subhaloes' DM profiles directly using the BCG+ICL profiles of galaxy clusters.

    \item The linear fit to the $\Sigma_{\rm cen-DM}-\Sigma_{\rm BCG+ICL}$ relationship involves some systematic deviations in the recovered $\Sigma_{\rm cen-DM}$ profile from the true one (Fig.~\ref{comparerectotruth}), especially by progressively overestimating the DM surface density at large radii, $r> 0.3r/R_{200}$ -- an effect mostly related to a combination of projection and the masking of satellites. Using our $\Sigma_{\rm cen-DM}-\Sigma_{\rm BCG+ICL}$ \textit{median} relation, the recovered DM surface density profile does not exhibit systematical deviations with radius and the uncertainty is well within 0.15 dex.
    
    \item The scatter in the scaling relation (i.e., the shaded region in each panel of Fig.~\ref{scaling}) exhibits a weak positive trend with DM mass (Fig.~\ref{scatcor}; Section~\ref{scatter}), and weak negative trends with formation redshift ($z_{\rm form}$), fraction of flux from BCG+ICL ($F_{\Delta,\,\rm BCG+ICL}$), and magnitude gap between the BCG and the $N$th most massive members ($\Delta m_{1N}$). This hints that some of the haloes that assembled later are still undergoing ICL and BCG formation, and possess greater DM content for the same BCG+ICL mass compared to early-assembled haloes, owing to the fact that DM loss from satellites
    generally commences earlier than stellar loss. Hence, part of the scatter in the $\Sigma_{\rm cen-DM}$--$\Sigma_{\rm BCG+ICL}$ scaling relations arises from varied formation times of our clusters. 
    This, however, does not account for \textit{most} of the scatter, which appears to be rather stochastic in origin.   
    
    \item As a consequence of this scatter, the halo parameters corresponding to the DM profiles recovered from the scaling relations -- either the median scalings or their linear approximations -- deviate from the true parameters for the central subhaloes' DM distributions (Fig.~\ref{dmvsrecfits}). Nonetheless, the relations between the central subhalo's DM parameters and those from the recovered DM profiles can be used to convert the latter into the former [equations~(\ref{c200cenvsrec})$-$(\ref{m500cenvsrecl})], resulting in concentrations with $\approx 0.1$~dex uncertainty, and the DM mass with an error of $\lesssim 0.07$~dex . 
    
    \item The central subhalo's DM parameters also present strong correlations with global photometric measurements (Fig.~\ref{dmvsphot}) which can be leveraged to estimate DM characteristics from optical observations.
    Namely, the concentration is correlated with the fraction of flux (or luminosity) contributed by 
    the BCG+ICL, and the mass correlates with the BCG+ICL flux. These relations offer an alterative route to derive the central subhalo's DM halo parameters from optical photometry. Similarly, the \textit{total} DM mass of the cluster correlates the best with the integrated/bolometric flux, and can be derived within $\approx 0.08$~dex of the true value.

    \item A direct comparison of the predictions for the central subhalo's DM parameters revealed
    that the global photometry typically results in predictions with less bias than those based on $\Sigma_{\rm cen-DM}$--$\Sigma_{\rm BCG+ICL}$ scalings (Fig.~\ref{comparepred}). The three methods (median scaling, linear scaling, and global photometry) are practically
    equivalent for deriving the $c_{200,\,\rm cen-DM}$ for the central subhaloes of clusters (Fig.~\ref{comparepred}), but can differ significantly for $M_{200,\,\rm cen-DM}$, $c_{500,\,\rm cen-DM}$,
    and $M_{500,\,\rm cen-DM}$. Notwithstanding, in cases with notable overmasking due to contaminations from foreground objects, it may not be feasible to carry out the global photometry be reliably, and the scaling relations should be used instead.

    \item Our tests on resolution and baryonic effects (Section~\ref{caveats}) showed that the former enhances the halo concentrations (Fig.~\ref{resconc}), whereas the latter reduces them (Fig.~\ref{barconc}). The impact is weaker on the concentrations from $\rho_{\rm cen-DM}$ profiles, 
    amounting to $\lesssim 4$ per cent in comparison to $\approx 3-8$ per cent for those from $\Sigma_{\rm cen-DM}$ profiles. For a given profile type, the changes differ between the two $R_\Delta$s
    by $\approx 1$ per cent.
    
\end{itemize}

Our work serves as a proof-of-concept for inferring DM radial distributions and DM halo characteristics from deep optical observations. Studies like this will aid in preparing for the upcoming surveys that are
poised to provide the ideal data sets for DM inferences, like the LSST at the Vera Rubin Observatory \citep{Ivezic2019}. Such relations could provide reasonable results 
for clusters at high redshifts even if they are based on low-$z$ snapshots \citep[e.g.][]{Asensio2025}. The next organic step for extending our analysis is to apply the results to WWFI images of real clusters, and compare the derived DM radial distributions and halo parameters to those from alternative methods -- accounting for the caveats as delineated in Section~\ref{recom}. We plan to carry this out in a follow-up paper.

We also would like to note that this study has utilised one of the many available simulation models that are well-suited for this kind of analysis; e.g. \eagle \citep{Schaye2015}, \simba \citep{Dave2019}. These simulations are broadly consistent with \tng but vary in their implementation of subgrid physics. Although this is thought to be mostly relevant for the circumgalactic gas and gas flows \citep{Crain2023}, it is worthwhile to perform studies 
similar to ours with other simulations to ascertain whether our results generally hold within different
models based on the $\Lambda$CDM framework.

\section*{Acknowledgements}
We would like to the thank the referee for a thorough review that has
aided in crucial improvements/additions to this paper. AM acknowledges support from Universidad Nacional Aut\'onoma de M\'exico Postdoctoral Program (POSDOC). The authors acknowledge support by the CONAHCyT grant `Ciencia de Frontera' G-543, and the DGAPA-PAPIIT grants IN106823, IN108323 and IN111825. DMT thanks CONAHCyT for a PhD fellowship.
The \tng simulations were conducted on the HazelHen Cray XC40 supercomputer at the High Performance Computing Center Stuttgart (HLRS) as part of project GCS-ILLU of the Gauss Centre for Supercomputing (GCS). This work has utilised the \textsc{\large python} packages \textsc{\large matplotlib} \citep{matplotlib}, \textsc{\large numpy} \citep{numpy}, \textsc{\large scipy} \citep{scipy}, \textsc{\large astropy} \citep{astropy}, and \textsc{\large photutils} \citep{photutils}. The paper has been typeset with Overleaf\footnote{\url{https://www.overleaf.com/}}. 

\section*{Data Availability}
The group catalogues and particle data for \tng [described in \citet{Nelson2019}] can be accessed at \url{https://www.tng-project.org/data/}. The median $\Sigma_{\rm cen-DM}$--$\Sigma_{\rm BCG+ICL}$ scaling relations and the corresponding percentiles (shown in Fig.~\ref{scaling}) are provided as supplementary material online. We also provide the median BCG+ICL profile of our clusters, and the median changes induced in their $\Sigma_{\rm cen-DM}$ profiles by resolution effects (Fig.~\ref{res}) and baryonification (Fig.~\ref{bar}). The mock optical images used in this work can be obtained via a reasonable request to the corresponding author, or generated using the \textsc{\large galaxev} pipeline at \url{https://github.com/vrodgom/galaxev_pipeline}.



\bibliographystyle{mnras}
\bibliography{paper} 



\appendix
\section{Profiles computed from masked map vs the central's map}\label{maskedvscencomp}
Observations of galaxy clusters are, by nature, limited to two dimensions, which poses a hurdle in accurate
characterisation of their three-dimensional mass distributions, and more so if one is trying
to distinguish the matter associated with the central/main subhalo. One of the approaches for the latter
involves masking the satellites in projected maps, as done in this paper (see Section~\ref{isolate}). Here, we examine whether the masking process results in $\Sigma$ profiles for our clusters similar to those for the centrals therein. We carry this out separately for the DM and the BCG+ICL  profiles.

\begin{figure}
    \centering
    \includegraphics[width=1\columnwidth]{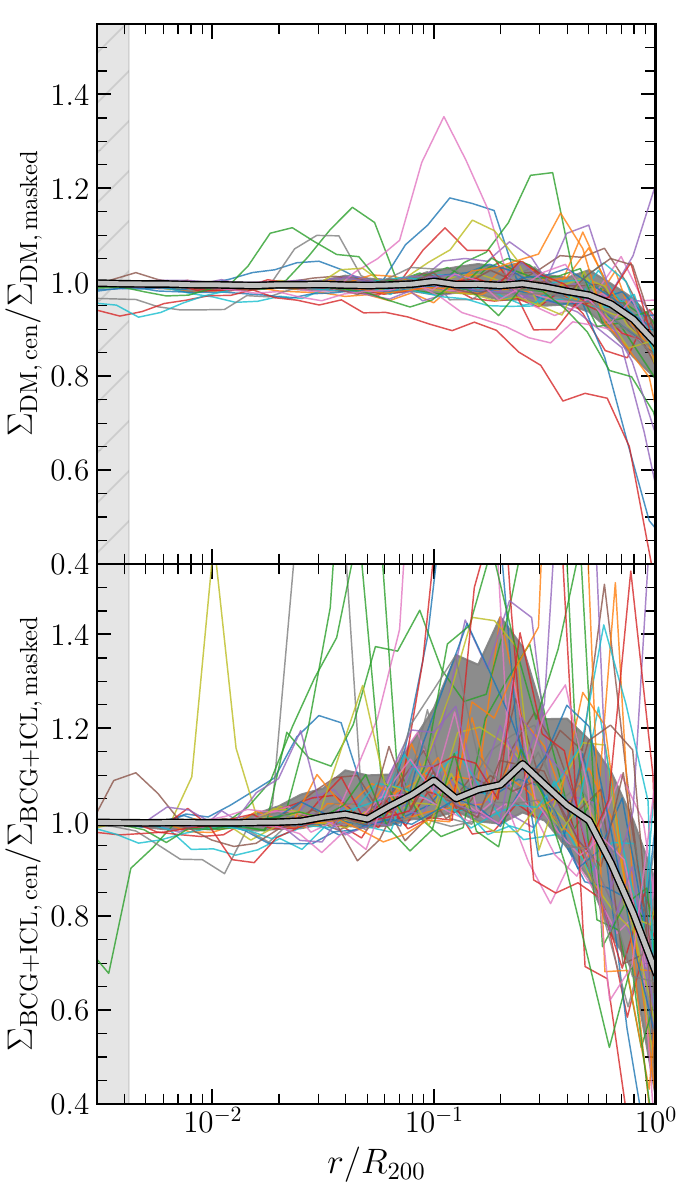}
    \caption{Comparisons of the projected DM ($\Sigma_{\rm DM,\,masked}$) and BCG+ICL ($\Sigma_{\rm BCG+ICL,\,masked}$) profiles constructed after masking the satellites (Section~\ref{isolate}) in the respective maps against the profiles corresponding to the centrals ($\Sigma_{\rm DM,\,cen}$ and $\Sigma_{\rm BCG+ICL,\,cen}$). The profiles extend out to $R_{200}$. The top panel shows the ratio of $\Sigma_{\rm DM,\,cen}$ profile to $\Sigma_{\rm DM,\,masked}$ profile for each cluster in our sample (differentiated using unique colours), and the bottom panel shows similar ratios for the BCG+ICL profiles. In each panel, the thick grey curve is the median ratio between the masked and centrals' profiles, and the shaded region
    surrounding it spans the 16th to 84th percentile. The vertical strip in the left depicts the regime where the profiles are not reliable due to two-body interactions between DM and baryonic
    particles (see Section~\ref{uniscale}). The DM profiles from masked maps are nearly the same as
    those for the central, but the BCG+ICL profiles are significantly disparate.}
    \label{maskedvscen}
\end{figure}

The results are shown in Fig.~\ref{maskedvscen}, where the top and bottom panels compare the DM and BCG+ICL
profiles, respectively. Each coloured curve shows the ratio of the profile obtained from the masked map
for the cluster to the one based on the central's map. The median ratio is shown using the thick grey curve,
and the shaded region around it spans 16th to 84th percentile. Additionally, we use the vertical
grey strip to demarcate the small radii where the profiles cannot be trusted, because they are potentially affected by two-body scattering between DM and baryonic particles (for details, see Section~\ref{uniscale}).

The top panel suggests that the central's DM profile is generally indistinguishable from that inferred from the masked DM map if $r\lesssim 0.3R_{200}$, but begins to deviate beyond this scale, exhibiting
progressively lower surface densities at larger radii. This implies a projection effect due to the masks and also that masking does not remove all the DM contained in satellites, likely because the masks are based on optical images and stars within satellites typically do not extend out as far as DM does. However, the discrepancies between the DM profiles tend to be $\lesssim 15$ per cent and are not particularly concerning with regard to our analysis. For instance, replacing the $\Sigma_{\rm DM,\,masked}$ profiles in Fig.~\ref{scaling} with $\Sigma_{\rm DM,\,cen}$ profiles only changes the linear fit parameters by $\lesssim 0.01$.

We now turn to the results for the BCG+ICL profiles in the bottom panel of Fig.~\ref{maskedvscen}. The differences are clearly more significant here, both in terms of the median and the scatter. 
At intermediate radii, $0.03\lesssim r/R_{\rm 200}\lesssim 0.5$, the masked surface brightness profiles are slightly lower, on average $<10$ per cent, than the centrals ones, while at radii $r>0.5R_{\rm 200}$ they are increasingly higher, by $\lesssim 40$ per cent or so, even if we only focus on the median ratio. 
Therefore, we conclude that, unlike DM, the BCG+ICL profiles from the masked images cannot be considered equivalent to those for the central.

\section{Radial variation of mass-to-light ratio}\label{masstolumvsrad}
Observation of a cluster provides the distribution of stellar \textit{light} in a certain wavelength band,
whereas simulations like TNG300 provide the stellar \textit{mass} distribution. One can assume
a direct correspondence between the two only if the mass-to-light ratio is constant across radii. However, there are indications that this may not be true. For a given mass, the emitted flux is a function of stellar metallicity and age, both of which are known to reduce with radius in galaxy clusters \citep{Montes2014,DeMaio2015,Montes2018,Gu2020}.

\begin{figure}
    \centering
    \includegraphics[width=1\columnwidth]{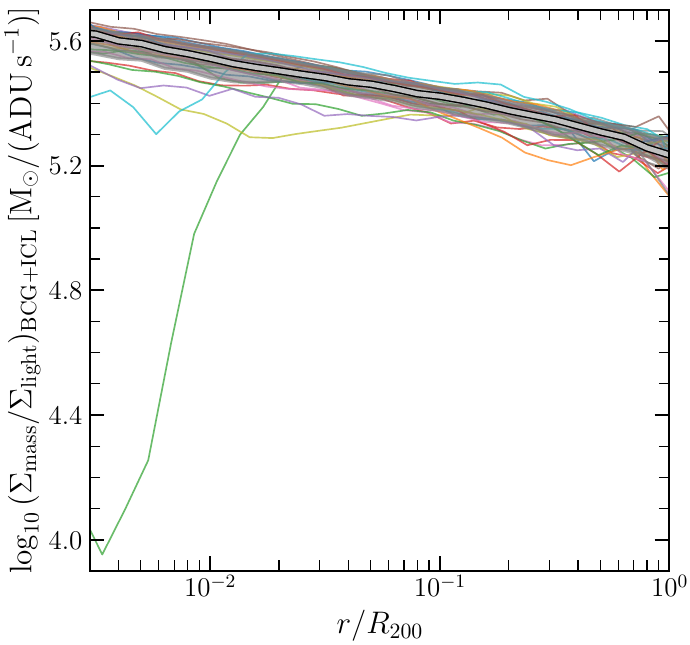}
    \caption{Ratios of projected stellar mass density ($\Sigma_{\rm mass}$) and the BCG+ICL surface brightness ($\Sigma_{\rm light}$) profiles for our TNG300 clusters. The stellar mass density profiles are obtained
    from the stellar mass maps after applying the same satellite masks as those used for the optical
    images. The thick grey curve is the median ratio and the shaded region
    surrounding it spans the 16th to 84th percentile. The mass-to-light ratio ($M/L$) reduces with cluster-centric radius, corresponding to $\approx 0.13$ dex decline in $M/L$ per one dex increase in radius.}
    \label{masstolight}
\end{figure}

For this reason, we investigate the mass-to-light ratio across radii within our TNG300 clusters. We 
generate the stellar mass maps that have the same resolution and field of view as the optical images, apply the same masks for satellite galaxies, and compute the stellar mass surface density profiles ($\Sigma_{\rm mass}$). In Fig.~\ref{masstolight}, we show the ratio of the $\Sigma_{\rm mass}$ to the surface brightness
profiles ($\Sigma_{\rm light}$) for the masked images, with the thick grey curve being the median
and the shaded region being the 16th to 84th percentile scatter. It is evident that the mass-to-light
ratio usually varies more than two folds ($\approx 0.4$~dex) within a cluster. For our sample,
this implies a $\approx 0.13$~dex decline for every one dex increase in normalised radius ($r/R_{200}$),
which is indeed expected given the typical metallicity and age gradients in clusters. This
clearly shows that one cannot simply take stellar mass maps of simulated clusters as proxies for light maps.

\section{Concentrations from $\Sigma_{\rm DM}$ and $\rho_{\rm DM}$ profiles of central subhaloes}\label{2dvs3d}
\begin{figure}
    \centering
    \includegraphics[width=0.9\columnwidth]{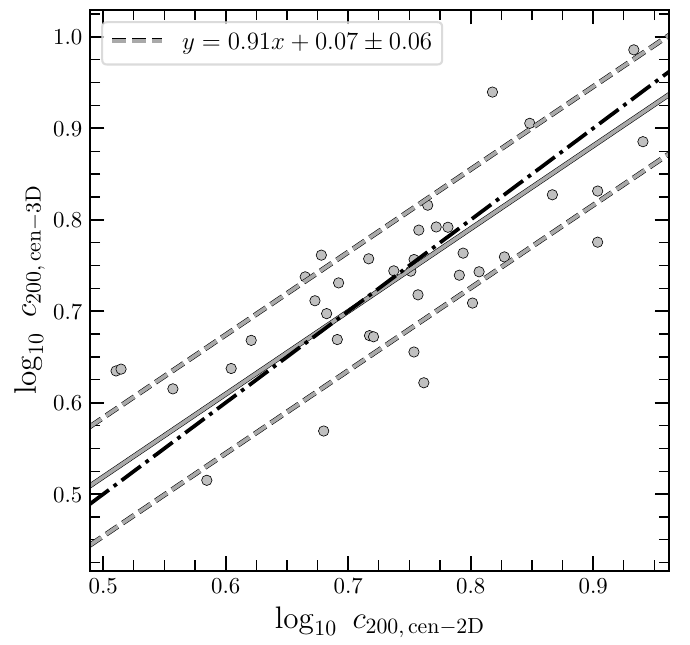}\quad
    \includegraphics[width=0.9\columnwidth]{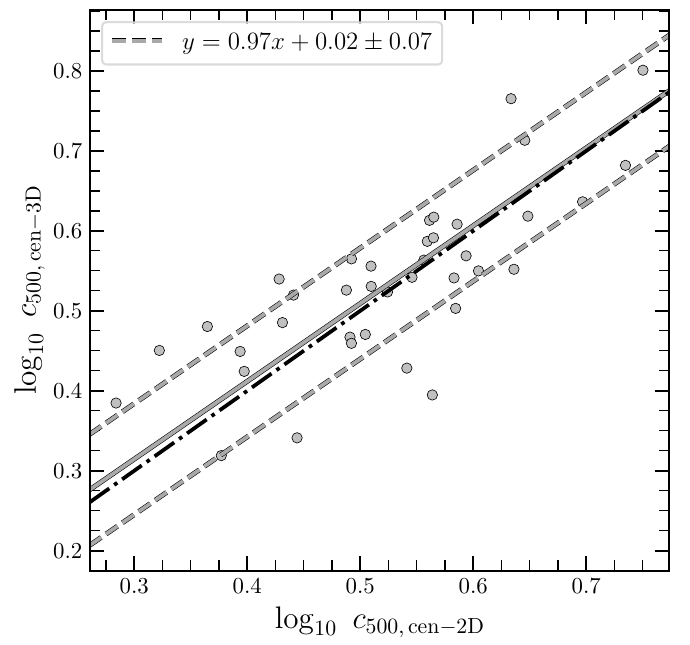}
    \caption{Comparison of DM concentrations of central subhaloes in our clusters derived from NFW fits to their 
    volumetric ($\rho_{\rm cen-DM}$) and surface density ($\Sigma_{\rm cen-DM}$) profiles. The top
    and bottom panels show the results for $\Delta=200$ and 500, respectively. The solid grey line
    is the best fit and the dashed grey lines show the $1$-$\sigma$ scatter. The dash-dotted
    line is the 1:1 relation.}
    \label{projandtrue}
\end{figure}

The NFW fits to the projected and three-dimensional density profiles would ideally yield the same concentrations,
but this is strictly valid only if the haloes exhibit perfect spherical symmetry. Here we examine
the impact of projection on the concentrations inferred for the central subhaloes in our sample. We
obtain the concentrations for the $\Sigma_{\rm cen-DM}$ and $\rho_{\rm cen-DM}$ profiles via the fitting approach described in Section~\ref{dmrec}, including the omission of small radii plagued by spurious heating. The results are shown in Fig.~\ref{projandtrue}, where the top panel
plots the concentrations derived from density profiles against those from the surface density profiles
extending out to $R_{200}$, and the bottom panel plots the quantities for the profiles spanning $r\lesssim R_{500}$.

It is clear that the two types of concentrations generally follow a relation close to 1:1 for $\Delta=200$,
but can deviate by $\lesssim 0.06$~dex. The concentrations for profiles extending out to $R_{500}$ show
a slope that is even closer to 1, but still exhibit a similar scatter ($\lesssim 0.07$~dex). Therefore,
projection effects cause a variation of $\lesssim 17$ per cent in the inferred DM concentrations
for the haloes in our sample.


\bsp	
\label{lastpage}
\end{document}